\documentclass[pra,twocolumn,superscriptaddress,showpacs]{revtex4}
\usepackage{graphicx,epsfig}
\begin{document}
\title{Effective Hamiltonians for periodically driven systems}

\author{Saar Rahav}
\affiliation{Department of Physics, Technion, Haifa 32000, Israel.}
\author{Ido Gilary}
\affiliation{Department of Chemistry, Technion, Haifa 32000, Israel.}
\author{Shmuel Fishman}
\affiliation{Department of Physics, Technion, Haifa 32000, Israel.}
\date{28 July 2003}

\begin{abstract}
The dynamics of classical and quantum systems which are driven
by a high frequency ($\omega$) field is investigated. For classical systems the 
motion is separated into a slow part and a fast part. The 
motion for the slow part is computed perturbatively 
in powers of $\omega^{-1}$ to order $\omega^{-4}$ and the corresponding
time independent Hamiltonian is calculated.
Such an effective Hamiltonian for the
corresponding quantum problem is computed to order
 $\omega^{-4}$ in a high frequency expansion.
 Its spectrum is the quasienergy
spectrum of the time dependent quantum system.
 The classical limit of this effective Hamiltonian is the classical effective time independent 
Hamiltonian. It is demonstrated that this effective Hamiltonian
gives the exact quasienergies and quasienergy states of some
simple examples as well as the lowest resonance of a non trivial model
for an atom trap. The theory that is developed in the paper is
useful for the analysis of atomic motion in atom traps of various shapes.
\end{abstract}

\pacs{42.50.Ct, 32.80.Lg, 03.65.Sq, 32.80.Pj} 

\maketitle

\section{Introduction}

The interaction of cold atoms with strong electromagnetic
fields results in many novel, interesting experimental observations~\cite{cornell02,tannudjibook2,faisal}.
The relevant systems are characterized by an extremely high degree
of control that enables one to explore various problems
of general physical interest. The response to a rapid
oscillating force is such issue and will be 
the subject of the present paper.

Recently, in a series of experiments, atomic billiards were
realized~\cite{raizen01,davidson01}. In these billiards
atoms were confined by a standing
wave of light to move in planes. The boundary of the billiards was 
generated by a laser beam, perpendicular to the plane of
motion. This beam rapidly traverses a closed curve, which
acts as the boundary of the billiard.
The boundary of the billiard is assumed to be
approximated by the time
average of this beam, and the force applied by the
boundary on the particles is approximately the mean force
applied by the beam.
 One expects that this approximation
is valid when the motion of the beam is fast relative to the typical
velocities of the atoms in the billiard. The billiards generated 
by the rapidly moving light beam motivated the
present work. The more general physical
problem, which is explored here, is the description of classical and quantum
dynamics in presence of fields that oscillate with high
frequency.

In traditional atomic physics one typically assumes that the fields which
affect the atoms have an amplitude which is constant in space
and is time independent. This assumption is justified since
the wavelength of the light field is much larger
than the size of the atom and the electronic (internal) degrees of freedom
react to the periodic change in the field much faster then the external ones (center of mass coordinate and momentum). 
The main subject of traditional atomic physics is the response
of the internal degrees of freedom to this field. Atomic spectroscopy
is the most spectacular result of this line of research.
The center of mass motion of the atom can be ignored in most
laboratory experiments that explore the dynamics of the 
internal degrees of freedom. 

For the field of atom optics
the effect of the internal degrees of freedom
on the center of mass motion is important, in particular near 
resonance of the external field with the internal motion (level spacing).
The force on the center of mass due to the internal degrees of freedom 
is given approximately by a dipole force~\cite{tannudjibook2}. The sign of this force depends on the sign of the detuning
of the light frequency from resonance (of the electronic levels).  
 The motion of the atoms is manipulated
 by fields with amplitudes which
vary spatially, resulting in a force on the center of mass of the atoms.
 In many cases the amplitude of the field
can be assumed static.
The atomic billiards described earlier consist of a time dependent
field which results of the moving laser beam. 
 Even at high frequencies of the motion of this beam one might
expect that this time dependence will have some dynamical consequences.
The question is most interesting when the wavelength of atoms
is of the order of the size of the billiards. 
In this work the effect of a laser on the center of mass
motion of the atoms will be modeled
by a time dependent potential.
For some situations of physical interest
 this simpler model should still
describe the dynamics in a high frequency field without
the need to specify the dynamics of the internal degrees of freedom or
the quantum aspects of the light field.
Therefore, in the present work the atoms are modeled by point particles
moving in a rapidly oscillating potential that varies in space.
This description is relevant for a wide class of light-atom
interactions and is not confined to models of billiards,
which motivated the present work.

The classical dynamics of particles influenced by a high frequency
field was studied in several contexts.
Kapitza investigated a classical pendulum with
a periodically moving point of suspension~\cite{kapitza}.
In this ``Kapitza's pendulum'' the motion can be separated
into a slow part and a fast part which consists of a rapid
motion around the slow part. The fast motion results 
in an effective potential for the slow motion.
In some range of parameters this pendulum performs
harmonic (slow) oscillations around the point where
it points upwards. This point is unstable in absence of the time
dependent perturbation. This phenomenon is called
``dynamical stabilization''. Later, Landau and Lifshitz generalized
this result for motion in presence of a rapid periodic force with
a spatially dependent amplitude~\cite{LL1} (see also \cite{percival}), and calculated the leading 
term in an expansion in powers of the inverse frequency. 

Dynamical stabilization is used 
to trap atoms in electromagnetic fields.
The most notable example is the Paul trap~\cite{paul}.
In this trap time dependent electric fields are used to localize
ions in the region where the field amplitude is minimal.
The fields are well approximated by restoring forces
which are linear in the distance from the equilibrium point.
The resulting Hamiltonian is that of a time dependent oscillator.
It is possible to find exact quantum mechanical solutions for this problem
that are based on the corresponding classical system. 
That is, the states are simply related to the ones of the Harmonic oscillator.
Therefore
the states of the motion in the Paul trap are known~\cite{glauber95,perelomov69,perelomovbook}.
It is of interest to find some of the states of problems of a more general 
nature, even if only approximately.

The work of Kapitza was first extended to quantum mechanical systems
in a pioneering paper by Grozdanov and Rakovi\'c~\cite{GR}. They introduced
a unitary gauge transformation resulting in an effective Hamiltonian
that describes the slow motion and demonstrated that its eigenvalues
are the quasienergies of the time dependent problem. The effective
Hamiltonian was calculated as an expansion in powers of the 
inverse frequency. In that paper the analysis is restricted
to a driving potential that has a particularly simple
time dependence. Moreover, the final results are restricted
to forces that are uniform in space, a situation natural
in standard spectroscopy, but too restrictive for the interesting
problems in atom optics. These restrictions are avoided in the present work.

 Other studies of quantum systems with periodic time dependent
fields were also published. Gavrila~\cite{gavrila96,gavrilabook}
developed a perturbation theory for the Floquet states and
the quasienergies in terms of the states of the time averaged
problem. The scattering from a periodically driven barrier
was studied by Vorobeichik et. al.~\cite{vorobeichik98}, by
Bagwell and Lake~\cite{bagwell92} and by Wagner~\cite{wagner97}, while
the quantum and classical dynamics of some one dimensional
systems were investigated by Henseler et. al.~\cite{henseler01}.
In the limit of high frequencies the systems behave as if
the particles were subject to an effective potential which is the
time average of the time dependent one. Fredholm theory was used by Georgeot and Prange~\cite{georgeot95} to study
quasiclassical scattering from various systems, including
a one dimensional periodically kicked potential.

Another approach to time dependent systems (not necessarily periodic)
is to use the Magnus expansion~\cite{magnus54} in order to
compute the propagator. Time periodic systems were used
as examples in order to check the convergence of
this expansion~\cite{maricq82,salzman87,fernandez90}.
For these time periodic systems the Magnus expansion
is of similar nature as the method presented here and
the differences are discussed in Sec.~\ref{theory}.

There are numerous other works regarding periodically driven systems.
Here we mention few of them. Of special physical interest is the ionization
of atoms by light, (see~\cite{gavbook} and references therein).
Some toy models for ionization that  consist of one dimensional time 
dependent $\delta$ functions were treated rigorously~\cite{costin}.
In particular, it was shown that typically there is full ionization,
namely, that at long times the probability to be at the bound state
(of the time averaged problem)
approaches zero. For some arrangements of the $\delta$ functions
 stable bound Floquet states exist. 
The transport  through driven mesoscopic
devices~\cite{li99} and in the presence of oscillating
fields~\cite{emman02} attracted some interest.

In the present work we study the dynamics of classical and quantum high-frequency driven
systems. 
The classical problem is discussed in Sec.~\ref{classical},
where the motion is separated into a ``slow'' and a ``fast'' part.
A systematic perturbation theory is developed for the motion
of the ``slow'' part. The equation of motion of the ``slow'' dynamics
is then computed to order $\omega^{-4}$,
which is an extension of the order $\omega^{-2}$
(resented in Ref.~\cite{LL1}). This slow motion is
shown to result from an effective Hamiltonian.
In Sec.~\ref{theory} an adaption of the Floquet theory 
to the problem is reviewed.
An effective (time independent) Hamiltonian operator is defined
following and generalizing~\cite{GR}.
The eigenvalues of this operator are the quasienergies of 
the system. This effective Hamiltonian is then computed perturbatively
(to order $\omega^{-4}$) in Sec.~\ref{highfreq}. 
The restrictions introduced in \cite{GR} are avoided and
consequently detailed expressions for the various terms of the effective
Hamiltonian are calculated explicitly.
The classical 
effective Hamiltonian of Sec.~\ref{classical} is found
to be the classical limit of this quantum effective Hamiltonian.
Two known exactly solvable simple examples
of the method are presented in Secs.~\ref{example1} and \ref{spin}.
In Sec.~\ref{gaussian} the scattering from a time dependent
potential is discussed. In particular, the resonances of the time dependent
problem are found to agree with those of the effective 
(time independent) Hamiltonian of Sec.~\ref{highfreq}. Finally, the results,
the implications and some related open problems are discussed in Sec.~\ref{discussion}.

\section{Classical motion in a high frequency potential}
\label{classical}

In this section the dynamics of a classical particle moving in one dimension
under the influence of a force which is periodic in time is studied.
Typically, solutions for time dependent problems can only be attained
numerically.
 However, when the period of the force
is small compared with other time scales of the problem it is possible
to separate the motion of the particle into ``slow'' and ``fast'' parts. This
simplification is due to the fact that the particle does not have the time
to react to the periodic force before this force changes its sign,
namely, the contribution of the periodic force to the acceleration in one period is negligible (compared to the contribution of the effective force in a sense
that will be specified in what follows).
Thus we will consider the limit of small periods (or large frequencies) of the driving field. 

The leading order (with respect to $1/\omega$) of the dynamics was computed by
Kapitza~\cite{kapitza} for the ``Kapitza's pendulum'', namely a pendulum
where the point of suspension is moved periodically. It turns out to be very general~\cite{LL1}. Here the next order 
is computed and it is demonstrated that the equation of motion of the slow
part of the dynamics can be derived from a time independent Hamiltonian. This Hamiltonian
will be computed explicitly to order $1/w^4$. Later, this Hamiltonian 
will be compared with an effective Hamiltonian which will be derived 
for the corresponding quantum problem.

The existence of such a Hamiltonian might seem to contradict the fact
that the time dependent dynamics do not possess a constant of motion.
Moreover, the classical motion may be chaotic. The existence of this effective time independent
Hamiltonian implies that a constant of motion exists
for the slow dynamics (it is just the effective Hamiltonian)
, and for a one dimensional
system the slow dynamics is integrable. To avoid confusion, it should
be emphasized that the effective Hamiltonian depends on a coordinate
which describes the ``slow'' part of the motion.
{\em This coordinate is not the location of the particle} (although they are almost identical at high frequencies).
The actual motion consists of rapid motion in the proximity of the trajectory
of the slow dynamics.
 The relation between the slow coordinate and the
coordinate of the particle is nonlinear
and extremely complicated as will be demonstrated in what follows. 
We will demonstrate that
an effective 
Hamiltonian for the ``slow'' motion may exist.

Newton's equation for the motion in the periodic field is given by
\begin{equation}
\label{newton}
m \frac{d^2x}{dt^2} = - V_0^\prime (x) - V_1^\prime (x, \omega t),
\end{equation}
where $V_1$ is a periodic function of $\omega t$ of period $2 \pi$ and 
its average over a period vanishes.
We denote derivatives with respect to coordinates by primes and with respect
to time by dots.
 This separation of the potential
to an average part, $V_0(x)$, and a periodic part with vanishing average, $V_1 (x,\omega t)$, is natural and will simplify the following calculations.
We look for a solution of the form
\begin{equation}
\label{separation}
x(t)=X(t)+\xi(X,\dot{X},\omega t)
\end{equation}
where $\dot{X}=\frac{dX}{dt}$ and
\begin{equation}
\label{xitauint}
\overline{\xi} \equiv \frac{1}{2 \pi} \int_0^{2 \pi} d \tau \xi(X,\dot{X}, \tau)=0.
\end{equation}
The bar denotes in this paper the time average over one period.
The fast part of the motion, that is nearly periodic in time, is denoted by $\xi$.
It will be shown later that it can be chosen to depend only on $X$ and
$\dot{X}$, but not on higher order time derivatives.
Since $X$ and $\dot{X}$ are slowly varying functions of the time
$t$, $\xi$ is {\em not} periodic in the time $t$, in spite of (\ref{xitauint}).
The coordinate $X$ describes the slow part of the motion and its equation of
motion will be computed in the following.
Our method of solution is to choose $\xi$
so that~(\ref{newton}) will lead to an equation for $X$ which is 
time independent. An exact solution using~(\ref{separation}) is too 
complicated to obtain. However, at high frequencies, one can determine
$\xi$ order by order in $1/\omega$. 
In order to separate terms in powers of the frequency it is convenient to 
introduce the new time variable $\tau \equiv \omega t$. Using
\begin{equation}
\frac{d \xi}{d t} = \omega \frac{\partial \xi}{\partial \tau} + \frac{\partial \xi}{\partial X} \dot X + \frac{\partial \xi}{\partial \dot{X}} \ddot{X}
\end{equation}
Newton's equation (\ref{newton}) is given by
\begin{widetext}
\begin{eqnarray}
\label{longn}
m \left( \ddot{X} + \omega^2 \frac{\partial^2 \xi}{\partial \tau^2} + 2 \omega \left[ \frac{\partial^2 \xi}{\partial X \partial \tau} \dot{X} +\frac{\partial^2 \xi}{\partial \dot{X} \partial \tau} \ddot{X} \right] + \frac{\partial \xi}{\partial X} \ddot{X} +  \frac{\partial \xi}{\partial \dot{X}} \dddot{X}
+ \frac{\partial^2 \xi}{\partial X^2} \dot{X}^2 + 2 \frac{\partial^2 \xi}{\partial X \partial \dot{X}} \dot{X} \ddot{X} +\frac{\partial^2 \xi}{\partial \dot{X}^2} \ddot{X}^2 \right) \nonumber \\  = -V_0^\prime (X+\xi) - V_1^\prime (X+\xi,\tau).
\end{eqnarray}
\end{widetext}
The variables $\tau$ and $t$ will be treated as independent variables.
This calculation is similar to the ones performed within the method of multiple time scales. Indeed, the result of the following
calculation is equivalent to the one obtained using the method of multiple time scales, 
as demonstrated in App.~\ref{multiple}.

In the limit of high frequencies $\xi$ is going to be small (of order $\omega^{-2}$) and therefore
it is convenient to expand $V_0 (X+\xi)$ and $V_1 (X+\xi,\tau)$ in powers
of $\xi$ (we assume that $V_0$ and $V_1$ are smooth functions the coordinate). 
Then $\xi$ is expanded in powers of $1/\omega$
\begin{equation}
\label{xiexpand}
\xi = \sum_{i=1}^{\infty} \frac{1}{\omega^i} \xi_i.
\end{equation} 
The $\xi_i$ are chosen so that the equation for $X$ that results from (\ref{longn}) does not depend on $\tau$. 

Before obtaining the slow equation of motion from
(\ref{longn}), order by order, there are two points regarding our method
of solution which should be discussed. 
First, we note that the fast part $\xi$ is expanded in powers of $1/\omega$ while
$X$ is not expanded, which seems to be inconsistent. 
One may also expand $X$ in powers of $1/\omega$ as $X=\sum_{i=0}^{\infty} \frac{1}{\omega^i} X_i$. When one does so 
the equation of motion for $X$ is replaced by a series of equations for $X_i$.
In this series of equations
each $X_i$ can be determined from the lower order terms $X_j$, where $j<i$.
This is the standard method of separation
of time scales and its application to the present problem
is demonstrated in App.~\ref{multiple}.
These equations are equivalent, in any order, to the equation of motion of the (unexpanded) $X$ which will be obtained in what follows.
At a given order $\omega^{-n}$ of the present calculation {\em all} 
contributions that are found by the method of separation
of time scales are included, but {\em some} of the higher
order terms are included as well.
Second, we note that while we assumed that $\xi$ depends only on $X$ and $\dot{X}$,
higher order derivatives of $X$ with respect to time appear in equation (\ref{longn}).
In the leading order in $1/\omega$, as will be demonstrated, one
can replace $\ddot{X}$ by $-\frac{1}{m} V_0^\prime (X)$. The error is of higher
order in $1/\omega$, leading to the correct contribution to $\xi_i$ at the order where $\ddot{X}$ appeared. Corrections of higher orders
of $1/\omega$ to $\xi_i$ result from the corrections of higher orders
to $\ddot{X}$. These corrections will affect the $\xi_j$ with $j>i$, since
these are chosen to cancel the $\tau$ dependence at any given order.
Higher order derivatives of $X$ can be found by repeated differentiation of $\ddot{X}$.
 This enables us to obtain an expression for $\xi$ that depends on $X$ and $\dot{X}$ but not on higher order derivatives
of $X$.

To proceed we gather all the terms in equation~(\ref{longn}), using~(\ref{xiexpand}), that are of the same order, say $\omega^{-n}$, and choose $\xi_{n+2}$
(which is still undetermined), so that 
the explicit $\tau$ dependence cancels.
In the leading order, ($\omega$), the only contribution is 
\begin{equation}
 \frac{\partial^2 \xi_1}{\partial \tau^2} = 0.
\end{equation}
Therefore we can choose
\begin{equation}
\xi_1=0.
\end{equation}
In the next order ($\omega^0$), we find the contributions
\begin{equation}
\label{classical0}
m \left( \ddot{X} + \frac{\partial^2 \xi_2}{\partial \tau^2} \right) = - V_0^\prime (X) - V_1^\prime (X, \tau).
\end{equation}
Our goal is to balance the $\tau$ dependence. 
To do this we have to solve
\begin{equation}
\label{before2}
\frac{\partial^2 \xi_2}{\partial \tau^2}= -\frac{1}{m} V_1^\prime (X, \tau),
\end{equation}
moreover, we also require that $\xi_2$ is periodic in $\tau$. The integral
over the right hand side (RHS) of equation (\ref{before2}) can have terms which
are time independent and thus $\xi_2$ can grow linearly in $\tau$.
To ensure that $\xi_2$ is small even at long times such secular terms must be
avoided.
This can be done by requiring that the time integral
has a vanishing average over a period. Let $f(x,\tau)$ be
any periodic function of $\tau$ with a vanishing average, $\overline{f} =0$.
Assume that the Fourier expansion of $f$ is given by $f=\sum_{n \ne 0} f_n e^{i n \tau}$,
then we define the following integral,
\begin{equation}
\label{defint}
 \int^\tau \left[ f \right] \equiv \sum_{n \ne 0} \frac{1}{in} f_n e^{i n \tau}.
\end{equation}
and its repeated application will be denoted by
\begin{equation}
 \int^{(2)\tau} \left[ f \right] = \int^\tau \left[ \int^\tau \left[ f \right] \right]
\end{equation}
and $j$ applications by
\begin{equation}
 \int^{(j)\tau} \left[ f \right] = \underbrace{\int^\tau \left[ \cdots \int^\tau \left[ f \right] \cdots \right]}_{\mbox{$j$ times}}.
\end{equation}
This definition, which is actually a specific choice of the
integration constant, is natural since it ensures that the result is periodic 
even after repeated integrations. It also helps to separate periodic terms
(with vanishing average) and secular terms (which will be time independent in
the current calculation).
Integration of (\ref{before2}) implies
\begin{equation}
\label{xi2}
\xi_2 = - \frac{1}{m} \int^{(2)\tau} \left[ V_1^\prime (X, \tau) \right].
\end{equation}
Note that we did not really find the general solution of~(\ref{classical0}), but rather chose $\xi$ so that it is satisfied. Substituting $\xi_2$ in equation (\ref{classical0}) gives the leading order equation for the slow coordinate $X$. The terms in this equation are just the
time independent terms which were not canceled by $\xi_2$,
\begin{equation}
\label{leadingslow}
m \ddot{X} = - V_0^\prime (X).
\end{equation} 

The contributions from~(\ref{longn}) at the next order, $\omega^{-1}$, are
\begin{equation}
m \left( \frac{ \partial^2 \xi_3}{\partial \tau^2} + 2 \dot{X} \frac{\partial^2 \xi_2}{\partial X \partial \tau} \right)=0
\end{equation}
which, with the help of~(\ref{xi2}), is satisfied by
\begin{equation}
\label{xi3}
\xi_3= \frac{2}{m} \dot{X} \int^{(3)\tau} \left[ V_1^{\prime \prime} (X,\tau) \right].
\end{equation}
\begin{widetext}
The terms of order $\omega^{-2}$ in equation~(\ref{longn}) are given by
\begin{equation}
m \left[ \frac{\partial^2 \xi_4}{\partial \tau^2} + 2 \left( \dot{X} \frac{\partial^2 \xi_3}{\partial X \partial \tau} + \ddot{X} \frac{\partial^2 \xi_3}{\partial \dot{X} \partial \tau}\right) + \ddot{X} \frac{\partial \xi_2}{\partial X}
+ \dot{X}^2 \frac{\partial^2 \xi_2}{\partial X^2} \right] = -V_0^{\prime \prime} (X) \xi_2 - V_1^{\prime \prime} (X,\tau) \xi_2.
\end{equation}
Substituting~(\ref{xi2}) and (\ref{xi3}) leads to
\begin{equation}
\label{b4xi4}
\frac{\partial^2 \xi_4}{\partial \tau^2} = \frac{V_0^{\prime \prime}}{m^2} \int^{(2)\tau} \left[ V_1^{\prime} \right] + \frac{V_1^{\prime \prime}}{m^2} \int^{(2)\tau} \left[ V_1^{\prime}\right] -\frac{3}{m} \dot{X}^2 \int^{(2)\tau} \left[ V_1^{(3)}(X,\tau) \right]- \frac{3 \ddot{X}}{m} \int^{(2)\tau} \left[ V_1^{\prime \prime} \right].
\end{equation}
\end{widetext}
Equation (\ref{b4xi4}) cannot be solved, if $\xi_4$ is required  to be periodic in $\tau$,
since the RHS has a non vanishing average which will lead to solutions that grow like $\tau^2$ (these are the secular 
solutions that one wishes to avoid when using multiple
time scales analysis). We will
choose $\xi_4$ so that it will balance the $\tau$ dependent part of (\ref{b4xi4})
and will be periodic in $\tau$.
The remaining $\tau$ independent terms in (\ref{b4xi4}) will be included in the equation of motion of the slow coordinate $X$.
Defining
\begin{equation}
f_1 (X,\tau) \equiv \frac{1}{m^2} V_1^{\prime \prime} \int^{(2)\tau} \left[
V_1^\prime\right]  - \frac{1}{m^2} \overline{V_1^{\prime \prime} \int^{(2)\tau} \left[
V_1^\prime \right]}
\end{equation}
and choosing
\begin{eqnarray}
\label{xi4}
\xi_4 & = & \frac{V_0^{\prime \prime}}{m^2} \int^{(4)\tau} \left[ V_1^\prime \right]+ \int^{(2)\tau} \left[  f_1\right] - \frac{3 \dot{X}^2}{m}  \int^{(4)\tau} \left[ V_1^{(3)} \right] \nonumber \\ & - & \frac{3 \ddot{X}}{m}  \int^{(4)\tau} \left[ V_1^{\prime \prime}\right]
\end{eqnarray}
balances all the $\tau$ dependent terms on the RHS of (\ref{b4xi4}) but leaves an extra term
$$ \frac{1}{m} \overline{V_1^{\prime \prime} \int^{(2)\tau} \left[ V_1^\prime \right]}$$
which is not balanced. This is actually a term of order $\omega^{-2}$
that is left on the RHS of~(\ref{longn}) when we substitute 
$\xi$ in~(\ref{longn}). The resulting equation for the slow motion is then
\begin{equation}
\label{slow2}
m \ddot{X} = -V_0^\prime (X) + \frac{1}{m \omega^2} \overline{V_1^{\prime \prime} \int^{(2)\tau} \left[ V_1^\prime \right]} + O(\omega^{-3}).
\end{equation}
This is the leading order correction due to the periodic potential $V_1$.
It was calculated before~\cite{kapitza,LL1}.
With the help of (\ref{leadingslow}) or of the leading term in (\ref{slow2})  $\ddot{X}$ can be eliminated from the expression
(\ref{xi4}) for $\xi_4$.
 This method allows us to compute
corrections order by order. We will continue the calculation up to order $\omega^{-4}$.

The next order is $\omega^{-3}$. We do not need to compute $\xi_5$ explicitly since it can only change the slow equation in order $\omega^{-5}$. To obtain the next correction to equation~(\ref{slow2}) one needs only the average over $\tau$ of the terms of order
$\omega^{-3}$. The reason is that $\xi_5$ will be chosen in such a way that it will cancel all the periodic terms with vanishing average.
This further simplifies the calculation since all the terms (except $m \ddot{X}$)
on the LHS of~(\ref{longn}) have vanishing average (over $\tau$) , thus only terms from the RHS can contribute
to the equation of the slow coordinate. In this order the contributions to the equation of the slow coordinate $X$ can result only from
\begin{equation}
-\overline{V_0^{\prime \prime} \xi_3}-\overline{V_1^{\prime \prime} \xi_3}.
\end{equation}
The first term will vanish since $V_0$ is $\tau$ independent and $\xi_3$ has
a vanishing average. The second term can be computed using (\ref{xi3})
\begin{eqnarray}
\overline{V_1^{\prime \prime} \xi_3} & = &\frac{2\dot{X}}{m} \overline{V_1^{\prime \prime} \int^{(3)\tau} \left[V_1^{\prime \prime} \right]} \nonumber \\ & = &
-\frac{2\dot{X}}{m} \overline{ \int^\tau \left[ V_1^{\prime \prime} \right] \int^{(2)\tau} \left[  V_1^{\prime \prime}\right]}=0.
\end{eqnarray}
In the last calculation we have used integration by parts and then the fact 
that the average of a derivative of a periodic function over a period must vanish. This leads to the conclusion that one can choose a periodic $\xi_5$ in such a way that
all $\tau$ dependent terms of order $\omega^{-3}$ in (\ref{longn}) are canceled.

We turn to the order $\omega^{-4}$ that is the last order that 
will be considered here. Again one can get the contributions to the
equation of $X$ by averaging
terms of this order in (\ref{longn}). 
The average over $\tau$ of the LHS vanishes and the contribution of the
terms on the RHS is
\begin{equation}
-\overline{V_0^{\prime \prime} \xi_4} -\frac{1}{2} \overline{V_0^{(3)} \xi_2^2}
-\overline{V_1^{\prime \prime} \xi_4} -\frac{1}{2} \overline{V_1^{(3)} \xi_2^2}.
\end{equation}
The first term will vanish but the other terms have a non vanishing average. Using~(\ref{xi2}), (\ref{xi4}) and  integration by parts (in the averages) yields
\begin{widetext}
\begin{eqnarray}
\label{cont4}
-\frac{1}{2} \overline{V_0^{(3)} \xi_2^2}
-\overline{V_1^{\prime \prime} \xi_4} -\frac{1}{2} \overline{V_1^{(3)} \xi_2^2}
=-\frac{1}{2m^2} V_0^{(3)} \overline{\left(\int^{(2)\tau} \left[ V_1^\prime \right] \right)^2} - \frac{1}{m^2} V_0^{\prime \prime} \overline{ \int^{(2)\tau} \left[ V_1^{\prime \prime} \right] \int^{(2)\tau} \left[ V_1^\prime \right]} 
-\frac{1}{2m^2} \overline{ V_1^{(3)}\left( \int^{(2)\tau} \left[ V_1^\prime \right] \right)^2} \nonumber \\ - \frac{1}{m^2} \overline{V_1^{\prime \prime} \int^{(2)\tau} \left[ V_1^{\prime \prime} \right] \int^{(2)\tau} \left[ V_1^\prime \right]} + \frac{3 \dot{X}^2}{m}\overline{ \int^{(2)\tau} \left[ V_1^{\prime \prime} \right] \int^{(2)\tau} \left[ V_1^{(3)} \right]}-  \frac{3 V_0^\prime}{m^2}\overline{ \left( \int^{(2)\tau} \left[ V_1^{\prime \prime} \right] \right)^2}.
\end{eqnarray}
In the last term the $\ddot{X}$ in $\xi_4$ was replaced by $-V_0^\prime/m$
resulting in errors that are of order $\omega^{-6}$ in the final result.
Eq. (\ref{cont4}) gives the $\omega^{-4}$ contribution to the equation for the slow coordinate
$X$. 

The equation  for $X$ to order $\omega^{-4}$ is obtained when $\xi$ is substituted into~(\ref{longn}) and the remaining terms are averaged over $\tau$
resulting in
\begin{eqnarray}
\label{slow4}
m \ddot{X} & =&  -V_0^\prime - \frac{1}{m\omega^2} \overline{\int^\tau \left[ V_1^\prime \right] \int^\tau \left[ V_1^{\prime\prime}\right]} -\frac{1}{2m^2 \omega^4} V_0^{(3)} \overline{\left( \int^{(2)\tau} \left[ V_1^\prime \right] \right)^2} 
 -  \frac{1}{m^2 \omega^4} V_0^{\prime \prime} \overline{ \int^{(2)\tau} \left[ V_1^{\prime \prime}\right] \int^{(2)\tau} \left[ V_1^\prime \right]} \nonumber \\ 
 & - & \frac{1}{2m^2 \omega^4} \overline{ V_1^{(3)}\left( \int^{(2)\tau} \left[ V_1^\prime \right] \right)^2} - \frac{1}{m^2 \omega^4} \overline{V_1^{\prime \prime} \int^{(2)\tau} \left[ V_1^{\prime \prime} \right] \int^{(2)\tau} \left[ V_1^\prime \right]} 
 +  \frac{3 \dot{X}^2}{m \omega^4}\overline{ \int^{(2)\tau} \left[ V_1^{\prime \prime}\right] \int^{(2)\tau} \left[ V_1^{(3)}\right]} \nonumber \\
&  - &  \frac{3 V_0^\prime}{m^2 \omega^4}\overline{\left( \int^{(2)\tau} \left[ V_1^{\prime \prime} \right]\right)^2 } + O(\omega^{-5}).
\end{eqnarray}
It is instructive to introduce the effective potential
\begin{equation}
\label{veff}
V_{eff} (X) \equiv V_0 + \frac{1}{2 m \omega^2} \overline{\left( \int^\tau \left[ V_1^\prime \right]\right)^2} + \frac{1}{2 m^2 \omega^4} \overline{V_1^{\prime\prime} \left( \int^{(2)\tau} \left[ V_1^\prime \right] \right)^2} + \frac{1}{2 m^2 \omega^4} V_0^{\prime \prime}\overline{ \left( \int^{(2)\tau} \left[ V_1^\prime \right] \right)^2}.
\end{equation}
\end{widetext}
Substituting (\ref{veff}) in (\ref{slow4}) results in the equation
of the slow motion
\begin{eqnarray}
\label{classicaleq}
m \ddot{X} =& - & V_{eff}^\prime +  \frac{3 \dot{X}^2}{m \omega^4}\overline{ \int^{(2)\tau} \left[ V_1^{\prime \prime} \right] \int^{(2)\tau} \left[ V_1^{(3)} \right]} \nonumber \\ & - &  \frac{3 V_0^\prime}{m^2 \omega^4}\overline{ \left( \int^{(2)\tau} \left[ V_1^{\prime \prime}\right]\right)^2 } + O(\omega^{-5}).
\end{eqnarray}
Given a solution of this $X(t)$, the solution for the original problem can be
easily obtained (to the appropriate order of $1/\omega$) since $\xi$ is known
in terms of $X$  (see~(\ref{xi2}), (\ref{xi3}) and (\ref{xi4})).
From these equations one sees that in the case where the oscillating force
$V_1^\prime$ is independent of position $X$ the fast coordinate
$\xi$ is independent of $X$ and $\dot{X}$ to the order $\omega^{-4}$
(note that in (\ref{xi4}) only the order $\omega^0$ of $X$ is
required see also (\ref{slowexpansion})). The final result of \cite{GR} is confined to
the case where $V_1^\prime$ is independent of $X$.
The equation (\ref{classicaleq}) can be derived
from the Hamiltonian
\begin{eqnarray}
\label{classicalheff}
 H_{eff} & = & \frac{P^2}{2 m} + V_{eff} (X) + \frac{3}{2 m^3 \omega^4} \overline{\left( \int^{(2)\tau} \left[ V_1^{\prime \prime} \right] \right)^2 } P^2 \nonumber \\ & + & O(\omega^{-5})
\end{eqnarray}
where $P$ is the momentum conjugate to $X$.

We have shown that using the natural separation of time scales
it is possible to separate the motion of a particle in a high frequency
periodic field into ``slow'' and ``fast'' parts.
The slow dynamics can be derived from an effective Hamiltonian which is time independent.
We turn to discuss the corresponding quantum problem.

\section{Floquet theory and the effective Hamiltonian}
\label{theory}

Consider a quantum system with a Hamiltonian that is periodic
in time, $\hat{H} (t+T)=\hat{H} (t)$. Such systems can be treated using
Floquet theory~\cite{zeldovich67,shirley65,sambe73,salzman74,gesztesy81}.
The symmetry with respect to discrete time translations
implies that the solutions of the Schr\"odinger equation
\begin{equation}
\label{schr}
i \hbar \frac{\partial}{\partial t} \psi = \hat{H} \psi
\end{equation}
are linear combinations of functions of the form
\begin{equation}
\label{quasistates}
\psi_{\lambda} = e^{-i\frac{\lambda t}{\hbar}} u_\lambda (x,\omega t)
\end{equation}
where
the $u_\lambda$ are periodic with respect to $\omega t$ with period $2 \pi$,
 that is $u_\lambda (x, \omega (t +T))=u_\lambda (x, \omega t )$
with $\omega=2 \pi/T$.
The states $u_\lambda$ are called the quasienergy or the Floquet states and $\lambda$ is 
referred to as the quasienergy (we will also call the states $\psi_\lambda$ quasienergy states). 
This is the content of the Bloch-Floquet theorem in time. 
The states $u_\lambda$ are the eigenstates of the Floquet Hamiltonian
\begin{equation}
\label{floquethamiltonian}
 \hat{{\cal H}}_F = -i \hbar \frac{\partial}{\partial t} + \hat{H}.
\end{equation}

The quasienergy (or Floquet) states have a 
natural separation into a ``slow'' part $e^{-i\frac{\lambda t}{\hbar}}$
(with the natural choice $0 \le \lambda/\hbar \le \omega$), 
which includes the information about the quasienergies,
and to a fast part $u_\lambda (x,\omega t)$ that depends only on
the ``fast'' time $\omega t$.
It is expected that one will be able 
to find an equation of motion for the slow part of the
dynamics as was done for classical systems in Sec.~\ref{classical}.
Such an equation will include information regarding the quasienergies
of the quantum system, and will be developed in what follows.
It establishes a natural link between the separation of the fast and the slow
motion in classical mechanics, which can be formalized by the theory
of separation of time scales, and Bloch-Floquet theory in quantum
mechanics.

It is known that one may write the propagator in the form~\cite{gesztesy81}
\begin{equation}
\label{prop1}
\hat{U} (t) = \hat{P} (t) e^{-i {t \hat{\cal G}}/{\hbar}}
\end{equation}
where $\hat{\cal G}$ is self-adjoint and $\hat{P}$ is unitary and periodic with
the period of the Hamiltonian.
The eigenvalues of $\hat{\cal G}$ are the quasienergies of the system provided
that the eigenstates of $\hat{\cal G}$ are in the domain of $\hat{H}(0)$.
Sometimes $\hat{\cal G}$ is called the quasienergy or Floquet operator.
The actual calculation of $\hat{\cal G}$ might be complicated.
Such an operator was calculated in~\cite{maricq82} and in~\cite{GR} by
introducing expansions for $\hat{P}$ and $\hat{\cal G}$. 
The result turns out to depend on the phase of the periodic part of the
Hamiltonian or on the initial time. (See for example \cite{maricq82}, equations (25) and (26) and \cite{GR}, equation (16)). Inspired by (\ref{schr})-(\ref{prop1}) an approach of a somewhat similar
spirit is used.

The goal is to find a unitary gauge transformation $e^{i \hat{F} (t)}$,
where $\hat{F} (t)$ is a hermitian operator (function of $\hat{x}$ and $\hat{p}$) defined at a certain time $t$, which is a periodic function of time
with the same period as $\hat{H}$, such that in the new gauge the
Hamiltonian in the
Schr\"odinger equation is {\em time independent}. 
Such a Hamiltonian was found by Grozdanov and Rakovi\'c~\cite{GR}
if the time dependent part is of the restricted form $V_{GR}=\tilde{V} (x) \sin (\omega t + \theta)$. It was analyzed with the further strong restriction,
that for one dimension takes the form $\frac{d \tilde{V}}{d x} = const$ 
(uniform force). In what follows a general analysis that is free of these
restrictions is presented.
Applying $e^{i\hat{F}}$ to
both sides of equation~(\ref{schr}) and adding $i \hbar \left( \frac{\partial}{\partial t}e^{i \hat{F}}\right) \psi$ to both sides leads to
\begin{equation}
i \hbar \frac{\partial}{\partial t} \left( e^{i \hat{F}} \psi \right) = e^{i \hat{F}} \hat{H} \psi + i \hbar \left( \frac{\partial}{\partial t} e^{i \hat{F}}\right) \psi.
\end{equation}
In terms of the functions in the new gauge, $\phi=e^{i \hat{F}} \psi$,
this equation is 
\begin{equation}
\label{schr2}
 i \hbar \frac{\partial}{\partial t} \phi= \hat{G} \phi,
\end{equation}
where the Hamiltonian is
\begin{equation}
\label{effect}
\hat{G} = e^{i \hat{F}} \hat{H} e^{-i \hat{F}} + i \hbar \left( \frac{\partial e^{i\hat{F}}}{\partial t} \right)e^{-i \hat{F}}.
\end{equation}
In the classical limit it reduces to
\begin{equation}
\label{classicalg}
 G=H- \hbar \frac{\partial F}{\partial t}.
\end{equation}
Therefore in the classical limit $-\hbar F$ is the generating function
of the canonical transformation corresponding to the unitary
transformation $e^{-i \hat{F}}$~\cite{goldstein}.

Let us assume that such an operator $\hat{F}$ exists so that $\hat{G}$ is
time independent.
Then the eigenfunctions of $\hat{G}$ are $v_\lambda (x)$,
their evolution takes the form
\begin{equation}
\phi_\lambda (t,x) = e^{-i \frac{\lambda t}{\hbar}} v_\lambda (x).
\end{equation}
These states, in the original gauge, correspond to
\begin{equation}
\label{back}
\psi_\lambda (t,x)= e^{-i \hat{F}} \phi_\lambda = e^{-i \frac{\lambda t}{\hbar}}  e^{-i \hat{F}} v_\lambda (x),
\end{equation}
since $\hat{F}$ does not include any time derivative. The function
$e^{-i \hat{F}} v_\lambda  $ is periodic in time with the period of $\hat{H}$
and therefore $\psi_\lambda$ is a Floquet state with quasienergy $\lambda$
(mod $\hbar \omega$). 
It should be compared with (\ref{quasistates}) with the identification $u_\lambda = e^{-i \hat{F}} v_\lambda$.
It is assumed that $e^{i \hat{F}}$ (and $e^{-i \hat{F}}$) are such that
they map the domain of $\hat{H} (t)$ into that of $\hat{G}$ and vice versa.
This may not be true in general, and 
one cannot exclude the possibility that
 examples, where only 
some of the quasienergies can be found using this method, exist.
For example, problems of this nature may occur if for a function $\psi$ in the Hilbert space of $\hat{H}$,
the function $e^{i \hat{F}} \psi$ is not in this space.
 The limitations
on the validity of the method should be subject to further
mathematical studies.

To emphasize the difference between the effective Hamiltonian $\hat{G}$
and $\hat{\cal G}$ given by equation~(\ref{prop1}) let us
write the propagator in terms of $\hat{F}$ and $\hat{G}$.
To propagate any state in time using $\hat{G}$ it has to be transformed to the
time independent gauge, then propagated and then transformed back.
This results in the propagator
\begin{equation}
\label{prop2}
\hat{U} (t) = e^{-i \hat{F} (t)} e^{-i \frac{t \hat{G}}{\hbar}} e^{i \hat{F} (0)}. 
\end{equation}
Since $\hat{G}$ and $\hat{F}$ do not commute, $\hat{G}$ generally differs
from $\hat{\cal G}$ of~(\ref{prop1}).

We note that an approximate solution of the time dependent problem in terms
of an expansion of $\hat{F}$ and $\hat{G}$ has some superior properties
compared to the more customary expansion of $\hat{\cal P}$ and $\hat{\cal G}$ of
(\ref{prop1}). For instance, if $\hat{F}$ is hermitian at any order than
$e^{i\hat{F}}$ is manifestly unitary while some care is needed to obtain
unitary approximations for $\hat{\cal P}$. In addition, $\hat{G}$ does not
depend on the phase of the time dependent field while $\hat{ \cal G}$
does depend on this phase (see \cite{GR,maricq82}).
Therefore in the present work a description in terms of $\hat{G}$ and $\hat{F}$
is used rather than one in terms of $\hat{\cal P}$ and $\hat{\cal G}$.

In the following section the derivation of $\hat{G}$ and $\hat{F}$ 
will be presented explicitly as an expansion
in powers of $1/\omega$. It will be shown that at high frequencies
$\hat{F}$ can be chosen to be small, of the order of $1/\omega$. In this limit
one can easily calculate matrix elements of an observable $\hat{O}$ between 
quasienergy (Floquet) states
using the eigenvalues and eigenstates of the effective Hamiltonian
$\hat{G}$
\begin{eqnarray}
\langle \psi_{\lambda_1} | \hat{O} | \psi_{\lambda_2} \rangle
 & = & \langle \phi_{\lambda_1} | e^{i \hat{F}} \hat{O} e^{-i \hat{F}} | \phi_{\lambda_2} \rangle \nonumber \\ & = & \langle \phi_{\lambda_1} | \hat{O} | \phi_{\lambda_2} \rangle + i\langle  \phi_{\lambda_1} | \left[ \hat{F},\hat{O}\right] | \phi_{\lambda_2} \rangle \nonumber \\ & - & \frac{1}{2}\langle \phi_{\lambda_1} | \left[ \hat{F}, \left[ \hat{F},\hat{O}\right]\right] | \phi_{\lambda_2} \rangle + \cdots .
\end{eqnarray}
The result is an effective expansion in powers of $1/\omega$.
It was obtained with the help of~(\ref{xepansion}) presented in App.~\ref{operator}.
Since observables have a meaningful classical limit $\hbar \rightarrow 0$
their expectation
 should reduce to the expansion in powers of $1/\omega$ for the
corresponding classical quantity as calculated in Sec.~\ref{classical}
and App.~\ref{multiple}. 
The expansion of $\hat{G}$ presented in the next section can be considered an
extension of the multiple scales analysis to quantum mechanics.
The effective Hamiltonian that will be obtained will be compared with the 
classical Hamiltonian for the slow motion that was computed in
Sec.~\ref{classical}.

\section{The effective Hamiltonian of quantum systems with a  high frequency potential}
\label{highfreq}

In Sec.~\ref{theory} we demonstrated that the quasienergies and Floquet states 
of a quantum system can be determined if one can find a gauge transformation
so that the Hamiltonian is time independent.
The transformation and the resulting effective Hamiltonian
are obtained here. Typically $\hat{F}$ and $\hat{G}$ cannot be computed exactly. For high frequencies
one can determine $\hat{F}$ and $\hat{G}$ order by order in $1/\omega$.
In the following we present a derivation of $\hat{F}$ and $\hat{G}$ 
accurate to order $1/\omega^4$.

We consider the Hamiltonian (that is more general than the one studied in~\cite{GR})
\begin{equation}
\label{quantumh}
\hat{H} = \frac{\hat{p}^2}{2 m} + \hat{V}_0 (x) +\hat{V}_1 (x, \omega t).
\end{equation}
This is the quantum system which corresponds to the classical system that
was discussed in Sec.~\ref{classical}. It should be noted that the method
which is described in the present section also applies to Hamiltonians
that differ from~(\ref{quantumh}), for example in presence
of magnetic fields and for spins (see Sec.~\ref{spin}).
We choose to examine the Hamiltonian~(\ref{quantumh}) since it is of
interest to compare the resulting effective Hamiltonian with its 
classical counterpart~(\ref{classicalheff}).
As mentioned in Sec.~\ref{theory} we are looking for a unitary transformation
$e^{i \hat{F}}$ so that the resulting Hamiltonian~(\ref{effect}) is time independent.
It is convenient to define $\tau= \omega t$,
since the Hamiltonian depends on time only through $\tau$. Using this definition
(\ref{effect}) is given by
\begin{equation}
\label{effect2}
\hat{G}= e^{i \hat{F}} \hat{H} e^{-i \hat{F}} + i \hbar \omega \left( \frac{\partial e^{i \hat{F}}}{\partial \tau} \right)e^{-i \hat{F}}.
\end{equation} 

At high frequencies $\hat{F}$ is assumed to be small, of the order of $1/\omega$.
An assumption that will be explicitly satisfied 
by the following calculation.  This enables
us to expand $\hat{G}$ and $\hat{F}$ in powers of $1/\omega$ and to choose $\hat{F}$ so that $\hat{G}$ is time independent in any given order. 
The expansions are given by
\begin{equation}
\label{expandg}
\hat{G} = \sum_{n=0}^{\infty} \frac{1}{\omega^n} \hat{G}_n
\end{equation}
and
\begin{equation}
\hat{F} = \sum_{n=1}^{\infty} \frac{1}{\omega^n} \hat{F}_n.
\end{equation}
The periodicity $\hat{F} (\tau + 2 \pi)= \hat{F} (\tau)$ is assumed.
The calculation is performed by computing $\hat{G}_l$ in terms of $\hat{F}_1, \cdots, \hat{F}_{l+1}$ and then choosing $\hat{F}_{l+1}$ so that $\hat{G}_l$
is time independent.
The terms in~(\ref{effect2}) are calculated with the help of the
operator expansions (presented in App.~\ref{operator}),
\begin{eqnarray}
\label{eifhemif}
 e^{i \hat{F}} \hat{H}  e^{-i \hat{F}}  & = & \hat{H} + i \left[ \hat{F}, \hat{H} \right] - \frac{1}{2!} \left[ \hat{F},\left[ \hat{F}, \hat{H} \right] \right] \nonumber \\ & - &
 \frac{i}{3!} \left[\hat{F}, \left[ \hat{F},\left[ \hat{F}, \hat{H} \right] \right] \right] + \cdots
\end{eqnarray} 
follows directly from (\ref{xepansion}) and
\begin{eqnarray}
\label{eifemif}
\left( \frac{\partial e^{i \hat{F}}}{\partial \tau} \right) e^{-i \hat{F}}  &= & i \frac{\partial \hat{F}}{ \partial \tau} - \frac{1}{2!} \left[ \hat{F},\frac{\partial \hat{F}}{ \partial \tau} \right] \nonumber \\ & - & \frac{i}{3!} \left[ \hat{F},\left[ \hat{F},\frac{\partial \hat{F}}{ \partial \tau} \right] \right] + \cdots 
\end{eqnarray}
is a result of (\ref{timeder2}) and (\ref{timecomm}).

In the leading order, $O(\omega^0)$, $\hat{G}_0$ is given by
\begin{equation}
\label{originalg0}
\hat{G}_0 = \frac{\hat{p}^2}{2 m} + \hat{V}_0 (x)+\hat{V}_1 (x,\tau) - \hbar \frac{\partial \hat{F}_1}{\partial \tau}. 
\end{equation}
The potentials $\hat{V}_0$ and $\hat{V}_1$ do not depend on $\hat{p}$.
To cancel any time dependence we choose
\begin{equation}
\label{f1}
\hat{F}_1 = \frac{1}{\hbar} \int^\tau \left[ \hat{V}_1 (x,\tau) \right].
\end{equation}
It is easily computed in the coordinate representation. Note that 
$\hat{F}_1$ is determined only up to a hermitian time independent operator.
It was assumed to vanish here.
Substituting~(\ref{f1}) in (\ref{originalg0}) leads to
\begin{equation}
\label{g0}
\hat{G}_0 = \frac{\hat{p}^2}{2 m}+ \hat{V}_0 (x).
\end{equation}
This is the leading order of the effective Hamiltonian. The dynamics
do not depend on the fast time dependent potential $\hat{V}_1$ as expected.
Corrections due to $\hat{V}_1$ will appear at higher orders in $1/\omega$. 

At order $1/\omega$ the effective Hamiltonian 
obtained from (\ref{effect2})-(\ref{eifemif}) is 
\begin{equation}
\label{originalg1}
\hat{G}_1 = i \left[ \hat{F}_1, \hat{H}\right] -\hbar \frac{\partial \hat{F}_2}{\partial \tau}- \frac{i \hbar}{2} \left[ \hat{F}_1, \frac{\partial \hat{F}_1}{\partial \tau} \right].
\end{equation}
Note that $\hat{F}_1$, given by~(\ref{f1}), depends only on the coordinate and therefore it commutes with its time derivative and also with $\hat{V}_0$. If a periodic $\hat{F}_2$ can be chosen so that
\begin{equation}
\label{df2dt}
\frac{\partial \hat{F}_2}{\partial \tau} = \frac{i}{\hbar}\left[ \hat{F}_1, \hat{H}\right]= \frac{i}{\hbar}\left[ \hat{F}_1, \frac{\hat{p}^2}{2m}\right], 
\end{equation}
then $\hat{G}_1$ vanishes.
Indeed, by choosing
\begin{equation}
\label{f2}
\hat{F}_2 = \frac{i}{2m} \int^{(2)\tau} \left[ V_1^{\prime \prime}\right] + \frac{i}{m}  \int^{(2)\tau} \left[ V_1^\prime \right] \frac{\partial}{\partial x}
\end{equation}
we obtain
\begin{equation}
\label{g1}
\hat{G}_1 = 0.
\end{equation}
We have presented $\hat{F}_2$ in the coordinate representation since
the simple dependence of $\hat{H}$ on momentum makes it the most
natural representation. We will use it also when calculating higher orders.

At the next order, $\omega^{-2}$, $\hat{G}_2$ found from  (\ref{effect2})-(\ref{eifemif}) is 
\begin{eqnarray}
\label{originalg2}
 \hat{G}_2 & = & i \left[ \hat{F}_2, \hat{H} \right] - \frac{1}{2} \left[\hat{F}_1 \left[ \hat{F}_1, \hat{H} \right] \right] - \hbar \frac{\partial \hat{F}_3}{\partial \tau} \nonumber \\ & - &\frac{i \hbar}{2} \left[ \hat{F}_1, \frac{\partial \hat{F}_2}{\partial \tau} \right] -\frac{i \hbar}{2} \left[ \hat{F}_2, \frac{\partial \hat{F}_1}{\partial \tau} \right] \nonumber \\ & + & \frac{\hbar}{6} \left[ \hat{F}_1 \left[ \hat{F}_1, \frac{\partial \hat{F}_1}{\partial \tau} \right] \right].
\end{eqnarray} 
Substituting $\hat{H}= \hat{G}_0 + \hbar \frac{\partial \hat{F}_1}{\partial \tau}$ and using (\ref{df2dt}) to eliminate the commutation relation $\left[ \hat{F}_1, \hat{H}\right]$ results in
\begin{equation}
\label{nextg2}
\hat{G}_2 = i \left[ \hat{F}_2 , \hat{G}_0 \right] - \hbar \frac{\partial \hat{F}_3}{\partial \tau} + \frac{i \hbar}{2} \left[ \hat{F}_2, \frac{\partial \hat{F}_1}{\partial \tau} \right].
\end{equation}
We can choose a periodic $\hat{F}_3$ in order to balance the time dependence
of $\hat{G}_2$. Note that $\hat{G}_2$ has some time independent part that
cannot be canceled by a periodic $\hat{F}_3$. Therefore we separate $\hat{G}_2$
into a $\tau$ independent part and a part that is periodic with vanishing average  and choose
$\hat{F}_3$ so that the latter vanishes (in (\ref{nextg2})).
For this purpose $\hat{F}_3$ must satisfy
\begin{equation}
\label{df3dt}
\frac{\partial \hat{F}_3}{\partial \tau} = \frac{i}{\hbar} \left[ \hat{F}_2, \hat{G}_0\right] + \frac{i}{2} \left( \left[ \hat{F}_2 , \frac{\partial \hat{F}_1}{\partial \tau}\right] - \overline{\left[ \hat{F}_2 , \frac{\partial \hat{F}_1}{\partial \tau}\right]} \right),
\end{equation}
where $\hat{G}_0$ is given by (\ref{g0}) and an average over a period is denoted by bar.
After some algebraic manipulations $\hat{F}_3$ is found to be
\begin{eqnarray}
\label{f3}
\hat{F}_3 & = & - \frac{\hbar}{m^2} \int^{(3)\tau} \left[ V_1^{\prime \prime} \right] \frac{\partial^2}{\partial x^2} - \frac{\hbar}{m^2} \int^{(3)\tau} \left[V_1^{(3)} \right] \frac{\partial}{\partial x} \nonumber \\ & - &\frac{\hbar}{4 m^2} \int^{(3)\tau}\left[ V_1^{(4)} \right] 
-  \frac{1}{m \hbar} V_0^\prime \int^{(3)\tau} \left[ V_1^\prime \right] \nonumber \\ & + & \frac{1}{2 m \hbar} \int^\tau \left[ {\cal P}_1 \right] + \hat{f}_3 (\hat{x},\hat{p})
\end{eqnarray}
where
\begin{eqnarray}
\label{defp1}
{\cal P}_1 (x,\tau) & \equiv & i m \hbar \left( \left[ \hat{F}_2,\frac{\partial \hat{F}_1}{\partial \tau}\right] - \overline{\left[ \hat{F}_2,\frac{\partial \hat{F}_1}{\partial \tau}\right]}\right)\nonumber \\ & = &\overline{V_1^\prime \int^{(2)\tau}\left[ V_1^\prime \right]} - V_1^\prime \int^{(2)\tau}\left[ V_1^\prime \right].
\end{eqnarray}
The constant of the integration over $\tau$ is the hermitian operator $\hat{f}_3$
that depends only on $\hat{x}$ and $\hat{p}$, and will be determined at
the next order.
Later we will use the freedom to choose $\hat{f}_3$ to cause $\hat{G}_3$ to have simple form.
Using (\ref{f3}) in (\ref{nextg2}) will cancel the time dependent terms resulting in:
\begin{eqnarray}
\label{g2}
\hat{G}_2 & = & \frac{i \hbar}{2} \overline{\left[ \hat{F}_2, \frac{\partial \hat{F}_1}{\partial \tau} \right]} = - \frac{1}{2m} \overline{ V_1^\prime \int^{(2)\tau}\left[ V_1^\prime \right]} \nonumber \\ & =  &\frac{1}{2m} \overline{\left( \int^\tau \left[V_1^\prime \right]\right)^2}
\end{eqnarray}
where we have used integration by parts.

The calculation of $\hat{G}_3$ and $\hat{G}_4$ can be performed
along similar lines. Since it is tedious it will be outlined in App.~\ref{highorders}. It starts from (\ref{originalg3}) and (\ref{originalg4})
that are obtained from  (\ref{effect2})-(\ref{eifemif}). Using the freedom
in the choice of $\hat{f}_3$ we choose it to satisfy (\ref{f3choice}), so that $\hat{G}_3$ vanishes. 
Then $\hat{F}_4$ is found to satisfy (\ref{df4dt}), that is integrated
to take the form (\ref{f4}). The time independent part of $\hat{F}_4$
is denoted by $\hat{f}_4$. Using the freedom of the choice of gauge,
we choose $\hat{f}_4$ so that in the classical limit the effective
Hamiltonian $\hat{G}$ reduces to its classical counterpart (\ref{classicalheff}).
This results in $\hat{G}_4$ of (\ref{g4}). In App.~\ref{highorders}
we calculated also $\hat{F}_4$ that was required for the calculation of
$\hat{G}_4$ but not $\hat{F}_5$ required for the calculation of $\hat{G}_5$. 

The freedom in the choice of gauge was used here and the time independent
parts of the $\hat{F}_i$ were chosen in a specific way. 
Generally this choice is arbitrary. In the present work, a choice was
made so that in the classical limit the effective Hamiltonian
reduces to the specific classical counterpart (\ref{classicalheff}), that resulted 
in a natural way within the derivation of Sec.~\ref{classical}.
In~\cite{GR}, on the other hand, the choice of the time independent
parts of the $\hat{F}_i$ is made so that the average of the fast
variables over a period reduces to the slow variables within an order $\omega^{-4}$
calculation. It is found there, that with this choice, the requirement can be 
satisfied only if the oscillating force is independent of position (in one dimension).

We have used perturbation theory (in $1/\omega$) to obtain a periodic
gauge transformation $e^{i \hat{F}}$ and an effective Hamiltonian $\hat{G}$
so that the quasienergies are the eigenvalues of $\hat{G}$.
Its eigenstates are related to the quasienergy states by (\ref{back}).
For a Hamiltonian of the form (\ref{quantumh}) this effective Hamiltonian
is given by equations (\ref{effect2}), (\ref{expandg}), (\ref{g0}),
(\ref{g1}), (\ref{g2}), (\ref{g3}) and (\ref{g4}). 
Collecting all contributions one finds,
\begin{widetext}
\begin{equation}
\label{finalg}
\hat{G} = \frac{\hat{p}^2}{2m} + \hat{V}_{eff} + \frac{1}{4 \omega^4} \left(\hat{p}^2 g(x) + 2 \hat{p} g(x) \hat{p}  +g(x) \hat{p}^2\right) + \frac{\hbar^2}{\omega^4} \hat{V}_q + O(\omega^{-5})
\end{equation}
where
\begin{equation}
\label{veff2}
V_{eff}(x) =  V_0 (x) + \frac{1}{2 m \omega^2}\overline{\left( \int^{\tau} \left[ V_1^\prime \right]\right)^2}  +\frac{1}{2m^2 \omega^4} V_0^{\prime \prime} \overline{\left( \int^{(2)\tau} \left[ V_1^\prime \right]\right)^2} + \frac{1}{2 m^2 \omega^4} \overline{V_1^{\prime \prime}\left( \int^{(2)\tau} \left[ V_1^\prime \right]\right)^2 }
\end{equation}
\end{widetext}
is the effective potential corresponding to (\ref{veff}),
\begin{equation}
\label{t}
g(x) = \frac{3}{2 m^3} \overline{ \left( \int^{(2)\tau} \left[ V_1^{\prime \prime} \right] \right)^2 },
\end{equation}
is the coefficient of $P^2$ in  (\ref{classicalheff}) while
\begin{equation}
\label{vq}
\hat{V}_q = \frac{1}{8 m^3} \overline{\left( \int^{(2)\tau} \left[ V_1^{(3)}\right]\right)^2}
\end{equation}
is a quantum correction to the classical Hamiltonian (its form obviously 
depends on the ordering of operators in (\ref{finalg})).

The effective Hamiltonian is the main result of this section. The classical limit of (\ref{finalg}) is the classical effective 
Hamiltonian (\ref{classicalheff}).
The freedom of gauge in the quantum problem was used, and $\hat{f}_3$ and $\hat{f}_4$ were chosen specifically to achieve
this.  We did not use the freedom of a canonical transformation in the classical calculation.
The specific canonical transformation from the Hamiltonian (\ref{quantumh}) to the Hamiltonian (\ref{classicalheff})
is generated by the classical limit of $- \hbar \hat{F}$ with the specific choice of the
time independent parts that was made in the present work.

The perturbation theory that was developed here enables one to calculate
not only the quasienergies but also the corresponding Floquet states.
If the eigenfunctions of $\hat{G}$ are known, then the quasienergy (or Floquet) states can be 
computed up to order $\omega^{-4}$ using equation (\ref{back})
with
\begin{eqnarray}
\label{finalf}
 \hat{F} & = & \frac{1}{\hbar \omega} \int^\tau \left[ V_1 \right] + \frac{i}{2 m \omega^2} \int^{(2)\tau} \left[ V_1^{\prime \prime} \right] \\ & + & \frac{i}{m \omega^2} \int^{(2)\tau} \left[ V_1^\prime \right] \frac{\partial}{\partial x} + \frac{1}{\omega^3} \hat{F}_3 + \frac{1}{\omega^4} \hat{F}_4 +O(\omega^{-5}) \nonumber
\end{eqnarray}
where $\hat{F}_3$ is given by (\ref{f3}) and (\ref{f3choice}) while
$\hat{F}_4$ is given by (\ref{f4}), (\ref{f4small}) and (\ref{functiong}).

\section{The harmonic oscillator driven by a periodic external force}
\label{example1}

In this section a 
 simple example that may help to clarify the meaning of the effective potential
and the gauge transformations used
in Sec.~\ref{highfreq} is discussed. 
It is the harmonic oscillator driven by a
force that is periodic in time.
It is defined by the Hamiltonian (\ref{quantumh}) with 
 $$ V_0 (x) =  \frac{1}{2} m \omega_0^2 x^2 $$
and
\begin{equation}
\label{forcedv0v1}
 V_1 (x,\tau) = {\cal E} x \cos (\tau).
\end{equation}
where $m$ is the mass of the particle, $\omega_0$ is the classical
frequency of the oscillator and ${\cal E}$ is the amplitude of the driving
force. We assume the non resonant case $\omega \ne \omega_0$.
For this simple system one is able to compute the quasienergies and the Floquet
states exactly. These were computed in several previous 
works~\cite{kerner58,perelomov69,perelomovbook,GR,breuer89,lefebvre97}.

The high-frequency perturbation theory of Sec.~\ref{highfreq} can be used
in order to calculate $\hat{F}$ and $\hat{G}$. The calculation is 
straightforward and substitution of (\ref{quantumh}) with (\ref{forcedv0v1}) in (\ref{finalg}) with (\ref{veff2})-(\ref{vq}) results in
\begin{equation}
\label{oscg}
\hat{G} =  - \frac{\hbar^2}{2m} \frac{\partial^2}{\partial x ^2} + \frac{1}{2} m \omega_0^2 x^2 + \frac{{\cal E}^2}{4 m \omega^2} + \frac{{\cal E}^2}{4 m \omega^2}\frac{\omega_0^2}{\omega^2} + O \left( \omega^{-5}\right)
\end{equation}
and substitution in (\ref{finalf}) yields
\begin{eqnarray}
\label{oscf}
\hat{F} & = &  \frac{{\cal E} x}{\hbar \omega} \sin \tau - \frac{{\cal E} i}{m \omega^2} \cos \tau \frac{\partial}{\partial x} + \frac{\omega_0^2 x {\cal E}}{\hbar \omega^3} \sin \tau  \nonumber \\ & + &  \frac{{\cal E}^2}{8 m \hbar \omega^3} \sin (2 \tau) - \frac{i \omega_0^2 {\cal E}}{m \omega^4} \cos \tau \frac{\partial}{\partial x}  + O \left( \omega^{-5}\right).
\end{eqnarray}
In this case both $\hat{G}$ and $\hat{F}$ have a simple form. 
The effective Hamiltonian $\hat{G}$ reduces to a Hamiltonian of a time independent Harmonic oscillator.
Equation~(\ref{oscf}) implies that the gauge transformation is of the linear form
(to order $1/\omega^4$)
\begin{equation}
\label{trialf}
\hat{F} = A(t) \hat{x} + B(t) + C(t) \hat{p}
\end{equation}
where $A,B,C$ are real periodic functions of time.
It turns out that (\ref{trialf}) is exact with
\begin{eqnarray}
\label{aandc}
A(t) & = & \nonumber \frac{{\cal E} \omega}{\hbar (\omega^2 - \omega_0^2)} \sin (\omega t) \nonumber \\
B(t) & = & \frac{{\cal E}^2}{ 8 \hbar m \omega (\omega^2 -\omega^2_0)} \sin (2 \omega t) \nonumber \\
C(t) & = & \frac{{\cal E}}{\hbar m (\omega^2 - \omega_0^2)} \cos (\omega t).
\end{eqnarray}
For this purpose one calculates $\frac{\partial \hat{F}}{\partial t}$,
and the commutators required for the expansions (\ref{eifhemif}) and (\ref{eifemif}).
These are very simple
since only $\left[ \hat{F} , \hat{H}\right]$ and $\left[ \hat{F}, \left[ \hat{F}, \hat{H} \right] \right]$
do not vanish. Requiring that the linear terms in $\hat{x}$ and $\hat{p}$
in the effective Hamiltonian $\hat{G}$ vanish determines $A(t)$ and $C(t)$ of (\ref{aandc}),
and the requirement that it is time independent results in $B(t)$ of (\ref{aandc}).
Finally the Hamiltonian $\hat{G}$ takes the form
\begin{equation}
\label{osceffective}
\hat{G} = \frac{\hat{p}^2}{2 m} + \frac{1}{2} m \omega_0^2 x^2 + \frac{{\cal E}^2}{4m (\omega^2 - \omega_0^2)}.
\end{equation}
Expansion of (\ref{osceffective}) in powers of $1/\omega$ results in (\ref{oscg})
while the expansion of (\ref{trialf}) with (\ref{aandc}) results
in (\ref{oscf}).
The eigenvalues of $\hat{G}$ are the quasienergies of (\ref{quantumh}) with (\ref{forcedv0v1}).
Since $\hat{G}$ is a harmonic oscillator Hamiltonian they are given by
\begin{equation}
\label{oscenergies}
E_n = \hbar \omega_0 ( n + \frac{1}{2} ) + \frac{{\cal E}^2}{4 m (\omega^2 -\omega_0^2)},
\end{equation}
with non negative integer $n$.
Note that these quasienergies are determined only up to integer multiples of
the period of $\hat{H}$, that is, they are given modulo $\hbar \omega$.
The Floquet states of this oscillator can also be computed. If $v_n (x)$
is the textbook eigenstate of the harmonic oscillator (\ref{osceffective}), with the energy $E_n$, 
than the corresponding Floquet state is $\psi_n (x,t) =  e^{-i \frac{E_n t}{\hbar}} e^{- i \hat{F}} v_n(x)$. 

The Floquet states ($u_\lambda$ of (\ref{quasistates})) are defined only up to powers of $e^{i n \omega t}$, that is, multiplying one of them by such a factor will also give a quasienergy state with the quasienergy shifted by $n \hbar \omega$.
It is possible to verify that these states are indeed eigenstates of the Floquet Hamiltonian (\ref{floquethamiltonian}),
and their eigenvalues are given by (\ref{oscenergies}).
One may require that all the quasienergies  are in the interval
$[ 0, \hbar \omega)$ and describe the dynamics by such
states. If $\omega/\omega_0$ is irrational almost any
value in this interval corresponds to a quasienergy. If $\omega/\omega_0$
is rational there is only a finite number of (infinitely degenerate)
quasienergies.

This harmonic oscillator driven by a periodic force serves as a simple example
for the method developed in Secs.~\ref{theory} and \ref{highfreq}. For this
simple example one can compute the quasienergies and Floquet states
of the system exactly~\cite{kerner58,perelomov69,GR,perelomovbook,breuer89,lefebvre97}. The perturbation theory described in Sec.~\ref{highfreq} leads to an expansion in powers of $1/\omega$. 
So far, we did not discuss the convergence of the series for the effective
Hamiltonian $\hat{G}$ and for $\hat{F}$. For the system discussed here both
series converge when $\omega>\omega_0$. The series can be resummed
also for $\omega<\omega_0$ and the result is valid for 
all $\omega \ne \omega_0$. In general the convergence properties
of the series for $\hat{F}$ and $\hat{G}$ are not known.
It is possible that the results of Sec.~\ref{highfreq} are meaningful also
for some cases where the series do not converge, if these can be resummed,
like the ones of the present example.

\section{Dynamics of spins in  time dependent magnetic fields}
\label{spin}

Another simple example that demonstrates the methods presented in this work
is a spin in a field, which is a combination of a static and
a periodic time dependent magnetic field. This example demonstrates that also
Hamiltonians that are not of the form $\hat{p}^2/{2m}+\hat{V}(x)$ can be 
treated in the way presented in Secs.~\ref{theory} and~\ref{highfreq}.  
The systems that are considered here consist of a spin in a constant magnetic field
combined with
 a perpendicular periodic field with linear or with circular polarization.
The Hamiltonian for the linearly polarized field is given by
\begin{equation}
\label{linear}
\hat{H}_l = - \omega_0 \hat{I}_z + \omega_1 \cos (\omega t) \hat{I}_x
\end{equation}
while for the circularly polarized field it is
\begin{equation}
\label{circular}
\hat{H}_c = - \omega_0 \hat{I}_z + \omega_1\left( \cos (\omega t) \hat{I}_x+\sin (\omega t) \hat{I}_y \right).
\end{equation}
For a spin in a circularly polarized field the problem was
solved exactly by Rabi~\cite{rabi37}. This system also appears
in textbooks as a paradigm of time dependent two level systems~\cite{tannoudji}. Our goal is to demonstrate that for this simple system, the quasienergies
can be computed exactly using the method presented in Secs.~\ref{theory} and~\ref{highfreq}. First we derive some results that are valid for any Hamiltonian
linear in the spin operators.

These spin problems turn out to be simple since the spin
operators have a closed algebra,
\begin{equation}
\label{scomm}
\begin{array}{ccc}
\left[ \hat{I}_x, \hat{I}_y \right]  =  i \hbar \hat{I}_z,\mbox{  } &
\left[ \hat{I}_y, \hat{I}_z \right]  =  i \hbar \hat{I}_x,\mbox{  } &
\left[ \hat{I}_z, \hat{I}_x \right]  =  i \hbar \hat{I}_y.
\end{array}
\end{equation}
The effective Hamiltonian (\ref{effect2}) is obtained with the help of the
expansions
in commutation relations (\ref{eifhemif}) and (\ref{eifemif}).
For a Hamiltonian and $\hat{F}$ that are linear in the spin operators 
these expansions can be summed. Consider a transformation generated by
\begin{equation}
\label{trialfspin}
\hat{F}= A(\tau) \hat{I}_x + B(\tau) \hat{I}_y+C(\tau) \hat{I}_z
\end{equation}
where $\tau=\omega t$ and $A$, $B$ and $C$ are real functions of time.
Let $\hat{Q}$ be an arbitrary operator that is linear in the $\hat{I}_i$. 
A straight forward calculation shows that
\begin{equation}
\label{doublecomm}
\left[ \hat{F}, \left[ \hat{F}, \left[ \hat{F}, \hat{Q} \right] \right] \right] = \alpha^2 \left[ \hat{F}, \hat{Q} \right]
\end{equation}
with
\begin{equation}
\label{defalpha}
\alpha = \hbar \sqrt{A^2 + B^2 + C^2}.
\end{equation}

Therefore for any Hamiltonian linear in the spin operators any commutation relation in (\ref{eifhemif}) can be reduced to $\left[ \hat{F} , \hat{H} \right]$
or to $\left[ \hat{F},\left[ \hat{F} , \hat{H} \right] \right] $ and 
the series is given by
\begin{widetext}
\begin{eqnarray}
 e^{i \hat{F}} \hat{H}  e^{-i \hat{F}} & = & \hat{H} -\frac{1}{2!} \left[ \hat{F}, \left[ \hat{F} , \hat{H} \right] \right] + \frac{1}{4!} \alpha^2 \left[ \hat{F}, \left[ \hat{F} , \hat{H} \right] \right] -\frac{1}{6!} \alpha^4 \left[ \hat{F}, \left[ \hat{F} , \hat{H} \right] \right] + \cdots \nonumber \\
 & + & i \left[ \hat{F}, \hat{H} \right] - \frac{i}{3!}  \alpha^2 \left[ \hat{F}, \hat{H} \right] + \frac{i}{5!}   \alpha^4 \left[ \hat{F}, \hat{H} \right] + \cdots \nonumber \\
& = & \hat{H} + \frac{\cos \alpha -1}{\alpha^2} \left[ \hat{F}, \left[ \hat{F} , \hat{H} \right] \right] + \frac{\sin \alpha}{\alpha} i \left[ \hat{F}, \hat{H} \right].  
\end{eqnarray}
The operator $\hat{F}$ of (\ref{trialfspin}) is linear in the spin operators and, therefore, such is also $\frac{\partial \hat{F}}{\partial \tau}$. In a similar manner (\ref{eifemif}) can be summed to
\begin{equation}
 \left( \frac{\partial \hat{F}}{\partial \tau} \right) e^{-i \hat{F}} = i \frac{\partial \hat{F}}{\partial \tau} + \frac{\sin \alpha - \alpha}{\alpha^3} i \left[ \hat{F}, \left[ \hat{F}, \frac{\partial \hat{F}}{\partial \tau} \right] \right] + \frac{\cos \alpha -1}{\alpha^2} \left[ \hat{F}, \frac{\partial \hat{F}}{\partial \tau} \right].
\end{equation}
The Hamiltonian in the new gauge is thus given by
\begin{equation}
\label{newspinh}
 \hat{G} =   \hat{H} + \frac{\cos \alpha -1}{\alpha^2} \left[ \hat{F}, \left[ \hat{F} , \hat{H} \right] \right] + \frac{\sin \alpha}{\alpha} i \left[ \hat{F}, \hat{H} \right] - \hbar \omega  \frac{\partial \hat{F}}{\partial \tau} 
+  \hbar \omega \frac{\cos \alpha -1}{\alpha^2} i \left[ \hat{F}, \frac{\partial \hat{F}}{\partial \tau} \right] - \hbar \omega \frac{\sin \alpha - \alpha}{\alpha^3}  \left[ \hat{F}, \left[ \hat{F}, \frac{\partial \hat{F}}{\partial \tau} \right] \right].
\end{equation}
\end{widetext}
The problem of finding the effective Hamiltonian is thus
reduced to finding three functions of time $A(\tau)$, $B(\tau)$ and $C(\tau)$
so that (\ref{newspinh}) is time independent.
Eq. (\ref{newspinh}) is valid for any Hamiltonian which is linear in the spin
operators.
Therefore, the problem is reduced to the solution of $3$ coupled nonlinear differential equations
that is a well defined mathematical problem.
 Generally this may be hard
to do since $\hat{G}$ is not linear in terms of these functions.
We turn now to examine the simplest case, that of a circularly polarized field~(\ref{circular}).

For the spin in a circularly polarized field a perturbative solution
in powers of $1/\omega$ for $\hat{F}$ and $\hat{G}$ can be found.
The computation is done exactly as the one of Sec.~\ref{highfreq}.
Thus only a brief
outline of the calculation is presented.
At the order $\omega^0$
\begin{eqnarray}
\label{startcirc}
 \hat{G}_0 & = & \hat{H} - \hbar \frac{\partial \hat{F}_1}{\partial \tau}\nonumber \\ & = & - \omega_0 \hat{I}_z + \omega_1 \left( \cos \tau \hat{I}_x + \sin \tau \hat{I}_y \right) - \hbar \frac{\partial \hat{F}_1}{\partial \tau},
\end{eqnarray}
and therefore
\begin{equation}
\hat{F}_1 = \frac{\omega_1}{\hbar} \left( \sin \tau \hat{I}_x - \cos \tau \hat{I}_y \right)
\end{equation}
and
\begin{equation}
\hat{G}_0 = - \omega_0 \hat{I}_z.
\end{equation}
Note that here $\left[ \hat{F}_1, \frac{\partial \hat{F}_1}{\partial \tau} \right] \ne 0$ which changes some of the expressions obtained in Sec.~\ref{highfreq}.

At order $\omega^{-1}$ a straight forward calculation leads to
\begin{equation}
 \hat{G}_1 = - \hbar \omega_0 \frac{\partial \hat{F}_1}{\partial \tau} -\hbar\frac{\partial \hat{F}_2}{\partial \tau}-\frac{\omega_1^2}{2} \hat{I}_z, 
\end{equation}
which results in
\begin{equation}
\hat{F}_2 = - \omega_0 \hat{F}_1= - \frac{\omega_0 \omega_1}{\hbar} \left( \sin \tau \hat{I}_x - \cos \tau \hat{I}_y \right)
\end{equation}
while
\begin{equation}
\hat{G}_1 = -\frac{\omega_1^2}{2} \hat{I}_z.
\end{equation}
A similar calculation at the next order  leads to
\begin{equation}
\hat{F}_3 = \left( \omega^2_0 - \frac{\omega_1^2}{3} \right) \hat{F}_1
\end{equation}
and to
\begin{equation}
\label{endcirc}
\hat{G}_2 = \frac{1}{2} \omega_0 \omega_1^2 \hat{I}_z.
\end{equation}

The expansions for $\hat{F}$ and $\hat{G}$ are obtained by collecting all 
the terms from Eqs. (\ref{startcirc})-(\ref{endcirc}). These
expansions are given by
\begin{equation}
\label{gcexp}
\hat{G}_c = \left( - \omega_0 - \frac{\omega_1^2}{2 \omega}+\frac{\omega_0 \omega_1^2}{2 \omega^2} + O(\omega^{-3}) \right) \hat{I}_z
\end{equation}
and by
\begin{eqnarray}
\label{fcexp}
\hat{F}_c & = & \frac{\omega_1}{\hbar \omega} \left( 1 - \frac{\omega_0}{\omega}+\frac{\omega_0^2}{\omega^2} - \frac{\omega_1^2}{3 \omega^2} + O(\omega^{-3}) \right)\nonumber \\ & \times & \left( \sin \tau \hat{I}_x - \cos \tau \hat{I}_y \right),
\end{eqnarray}
where the subscript c denotes that this result is obtained for the circularly polarized
field.

An examination of Eq. (\ref{fcexp}) suggests that $\hat{F}_c$
may have the exact form
\begin{equation}
\label{fcguess}
\hat{F}_c = \frac{\alpha (\omega)}{\hbar} \left( \sin \tau \hat{I}_x - \cos \tau \hat{I}_y \right).
\end{equation} 
It turns out that $\hat{F}_c$ of this form leads to a time independent
Hamiltonian if $\alpha$ is chosen appropriately.
Substituting $\hat{F}_c$ and $\hat{H}=\hat{H}_c$ of (\ref{circular}) in (\ref{newspinh})
leads to
\begin{eqnarray}
\label{newgaugec}
\hat{G}_c & = & \left( \omega_1 \cos \alpha - \omega_0 \sin \alpha - \omega \sin \alpha \right) \left( \cos \tau \hat{I}_x +\sin \tau \hat{I}_y \right) \nonumber \\
& + & \left( - \omega_0 \cos \alpha - \omega_1 \sin \alpha+ \omega \cos \alpha + \omega \right) \hat{I}_z.
\end{eqnarray}
The Hamiltonian $\hat{G}_c$ is time independent if
\begin{equation}
\omega_1 \cos \alpha - \omega_0 \sin \alpha - \omega \sin \alpha =0.
\end{equation}
Solving for $\alpha$ the Hamiltonian in the new gauge, (\ref{newgaugec}) reduces to
\begin{equation}
\label{effectivec}
\hat{G}_c = \left( \omega - \sqrt{\omega_1^2 +(\omega_0+\omega)^2} \right) \hat{I}_z,
\end{equation}
and is time independent.
The quasienergies of the spin in a circularly polarized field are the eigenvalues
of (\ref{effectivec})). They are
given by
\begin{equation}
\label{spinquasi}
 E_s = \left( \omega - \sqrt{\omega_1^2 +(\omega_0+\omega)^2} \right) \hbar s
\end{equation}
where $s=-S,-S+1,\cdots, +S$  ($S$ is the magnitude of the spin).

For spin $S=1/2$ not only the quasienergies but also the quasienergy states 
can be computed rather easily. The spin operator can be represented
by the Pauli matrices
\begin{equation}
\begin{array}{ccc}
 \hat{I}_x = \frac{\hbar}{2} \left( \begin{array}{cc} 0 & 1 \\ 1 & 0 \end{array} \right), \! &
 \hat{I}_y =  \frac{\hbar}{2} \left( \begin{array}{cc} 0 & -i \\ i & 0 \end{array} \right), \! &
\hat{I}_z = \frac{\hbar}{2} \left( \begin{array}{cc} 1 & 0 \\ 0 & -1 \end{array} \right). 
\end{array}
\end{equation}
The Hamiltonian is then given by
\begin{equation}
\label{matrixhc}
\hat{H}_c = \frac{\hbar}{2} \left( \begin{array}{cc} -\omega_0 & \omega_1 e^{-i\omega t} \\ \omega_1 e^{i \omega t} & \omega_0 \end{array} \right).
\end{equation}
The unitary transformation $\hat{U}_c=e^{-i \hat{F}_c}$ which transforms the
eigenstates of the effective Hamiltonian (\ref{effectivec})
to the quasienergy states of (\ref{matrixhc}) can be obtained
by calculating the various powers of $\hat{F}_c$.  For $S=1/2$,
\begin{equation}
\hat{F}_c = \frac{\alpha}{2i} \left( \begin{array}{cc} 0 & - e^{-i\omega t} \\  e^{i \omega t} & 0 \end{array} \right).
\end{equation}
Since $\hat{G}_c$ is proportional to $\hat{I}_z$ its eigenstates are the
eigenstates of $\hat{I}_z$. Thus, the quasienergy states of the Hamiltonian  (\ref{matrixhc}), corresponding to quasienergies 
\begin{equation}
E_\pm = \pm \frac{\hbar}{2} \left( \omega - \sqrt{\omega_1^2 +(\omega_0+\omega)^2} \right)
\end{equation}
 are: 
\begin{equation}
u_+ = \hat{U}_c \left( \begin{array}{c} 1 \\ 0 \end{array} \right) = \left(  \begin{array}{c} \cos \frac{\alpha}{2} \\ - e^{i \omega t} \sin \frac{\alpha}{2} \end{array} \right)
\end{equation}
and
\begin{equation}
u_- = \hat{U}_c \left( \begin{array}{c} 0 \\ 1 \end{array} \right) = \left(  \begin{array}{c} e^{- i \omega t} \sin \frac{\alpha}{2} \\  \cos \frac{\alpha}{2} \end{array} \right).
\end{equation}
This is
exactly the problem that was solved by Rabi~\cite{rabi37} and is discussed in textbooks~\cite{tannoudji}. The physical quantity of interest
is typically the amount of spins that flip (if all spins are polarized initially), rather than the Floquet states.
We note that the term $\sqrt{\omega_1^2+(\omega_0+\omega)^2}$  in the expression for the quasienergies is the Rabi frequency. It is the frequency
of oscillations of these ``spin flips''. 

For the spin in a linearly polarized field a perturbative computation of 
$\hat{G}$ leads to
\begin{equation}
\label{approxspinl}
\hat{G}_l = \left( - \omega_0 + \frac{\omega_0 \omega_1^2}{4 \omega^2} + \frac{\omega_0^3 \omega_1^2}{4 \omega^4} - \frac{\omega_0 \omega_1^4}{64 \omega^4} + O(\omega^{-5}) \right) \hat{I}_z.
\end{equation}
To the best of our knowledge an exact expression for the quasienergies
of this system is not known. If one substitutes $\hat{F}$ of the form
(\ref{trialfspin}) in (\ref{newspinh}), 
the problem
of finding the effective time independent Hamiltonian reduces to the problem
of finding the three functions of time, $A(\tau)$, $B(\tau)$ and $C(\tau)$, so that the new Hamiltonian is time independent.
These satisfy first order nonlinear differential
equations. Typically solutions to such equations exist but it is
not easy to find them explicitly. It is possible to choose parameters
so that also the exact $\hat{G}_c$ is proportional to $\hat{I}_z$.
The approximate effective Hamiltonian (\ref{approxspinl}) can be
compared to previously published results. While we are not aware of 
any $1/\omega$ expansion for the quasienergies, some expansions
in the strength of the time dependent field have been published.
If one examines, for instance, the expansion given by Eqs. (2.10)
and App. A of~\cite{barone77} (which is valid for spin $S=1/2$)
and expands it in powers of $1/\omega$ one obtains the quasienergies
of (\ref{approxspinl}).

In this section we have studied some problems involving spins in crossed
constant and time dependent magnetic fields. We have shown that the
perturbation theory presented in Sec.~\ref{highfreq} can be used
for such systems.
For a circularly polarized field we were able to compute the
quasienergies exactly, in agreement with
previously published results.

\section{Scattering from an oscillating Gaussian potential}
\label{gaussian}

The systems considered in Secs.~\ref{example1} and \ref{spin} are simple,
in the sense that the spectrum of the effective Hamiltonian $\hat{G}$ is
discrete and simply related to the one of the
time independent part of the original Hamiltonian $\hat{H}_0$. Moreover,
for these examples also the eigenfunctions of these Hamiltonians
are simply related.
It is of interest to examine examples that are more
complicated and where such simple relations cannot be found.
In this section we examine such a system, the oscillating Gaussian,
where an additional difference is that the spectrum is continuous
and one is interested in scattering states.

Consider a system which consists of a particle that interacts
with an oscillatory Gaussian potential. The Hamiltonian is
given by
\begin{equation}
\label{oscgauss}
\hat{H} = \frac{\hat{p}^2}{2m} + \gamma e^{- \beta x^2} \cos (\omega t).
\end{equation}
The system is of interest for two reasons. First, when $x \rightarrow \infty$
the potential vanishes and therefore one expects to find 
scattering quasienergy states. Second, the average of the potential
vanishes, namely $V_0(x)=0$, consequently any interesting effect is due to the rapidly oscillating  potential.
This system describes trapping by an oscillating field, a phenomenon that is of
physical interest. The physical properties of this system and 
the numerical methods used to analyze it are discussed elsewhere~\cite{ido}.
Here we only state briefly the results that are related to
the properties of the effective Hamiltonian.

The effective Hamiltonian~(\ref{finalg}), that corresponds to (\ref{oscgauss}), is 
\begin{widetext}
\begin{eqnarray}
\label{order4gauss}
\hat{G} & = & \frac{\hat{p}^2}{2m} + \frac{\beta^2 \gamma^2 x^2}{m \omega^2} e^{- 2 \beta x^2} + \frac{3 \beta^2 \gamma^2}{m^3 \omega^4} \left( 1 - 2 \beta x^2 \right)^2 e^{-2 \beta x^2} \hat{p}^2 -   \frac{12 i \hbar \gamma^2 \beta^2 x}{m^3 \omega^4} \left( 2 \beta x^2 -1 \right) \left(3 - 2 \beta x^2\right) e^{-2 \beta x^2} \hat{p} \nonumber \\ & - & \frac{\hbar^2 \beta^3 \gamma^2}{m^3 \omega^4} \left( -9 +99\beta x^2 -114 \beta^2 x^4 + 44 \beta^3 x^6\right) e^{-2 \beta x^2} + O(\omega^{-5}). 
\end{eqnarray}
\end{widetext}
We examine separately the leading correction due to
the oscillating field, that is given by the Hamiltonian
\begin{equation}
\label{effectgauss}
\hat{G}^{(2)} \equiv \hat{G}_0 + \hat{G}_1 + \hat{G}_2 =  \frac{\hat{p}^2}{2m} + \frac{\beta^2 \gamma^2 x^2}{m \omega^2} e^{- 2 \beta x^2},
\end{equation}
where the error is of order $\omega^{-4}$.
It has a simple physical meaning as the  Hamiltonian of a double barrier potential and 
its spectrum is continuous.

Since the effective potential of Eq. (\ref{effectgauss}) is a double barrier
one expects to find that this system exhibits resonances.
 These resonances
describe long lived unstable states. 
Each resonance is characterized by a complex energy $E-i\Gamma/2$.
The real part $E$ is the location of the resonance while $\Gamma$
is the width which is inversely proportional to the lifetime.
For a review on relevant properties
of resonances and useful methods to compute them see~\cite{nimrev}.

For any resonance of (\ref{order4gauss}) and (\ref{effectgauss})  it is natural to look
for the corresponding resonance of the time dependent original Hamiltonian~(\ref{oscgauss}).
More precisely, one looks for the resonances of the Floquet 
Hamiltonian (\ref{floquethamiltonian}) with $\hat{H}$ of (\ref{oscgauss}). This is done
numerically using a combination of the ($t,t'$) method
and complex scaling~\cite{nimrev}.

\begin{figure}[htb]
\includegraphics[width=7cm]{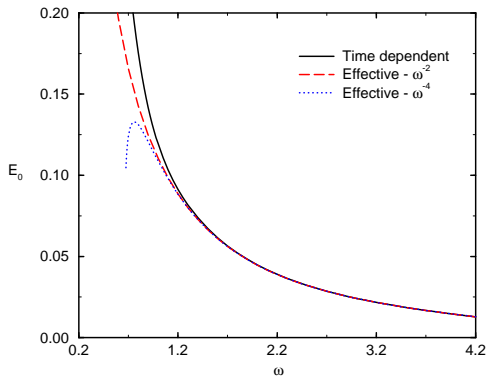}
\caption{The energy $E_0$ of the lowest quasienergy resonance of the Hamiltonian ({\protect{\ref{oscgauss}}}) as a function of the driving frequency (solid line), compared to the lowest resonance of the effective Hamiltonians ({\protect{\ref{effectgauss}}})--dashed line, and ({\protect{\ref{order4gauss}}})--dotted line, for $\gamma=9$ and $\beta=0.02$. The ``atomic units'' $m=\hbar=e=1$ are used here.\label{energy}} 
\end{figure}

\begin{figure}[htb]
\includegraphics[width=7cm]{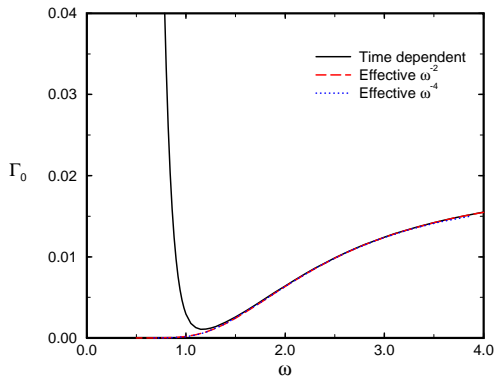}
\caption{Same as Fig.~{\protect{\ref{energy}}} for $\Gamma_0$, the width of the lowest resonance. \label{width}}
\end{figure}

The energy $E_{0}$ and the width $\Gamma_{0}$ of the lowest (corresponding
to the smallest real part $E_0$) quasienergy-resonance
of (\ref{oscgauss}) are compared with the lowest resonance of
the effective Hamiltonians (\ref{order4gauss}) and  (\ref{effectgauss}) in Figs.~\ref{energy} and 
\ref{width}.
It is clear that for large frequencies there is excellent agreement
between the resonance of the time dependent Hamiltonian~(\ref{oscgauss})
and the ones of the effective Hamiltonians~ (\ref{order4gauss}) and (\ref{effectgauss}).
At low frequencies the location and width of the exact resonance
differ from those of the effective Hamiltonian. 
The deviation for the order $\omega^{-4}$ Hamiltonian (\ref{order4gauss})
is large indicating that the expansion is asymptotic. This
is expected since the perturbation theory developed in Sec.~\ref{highfreq}
assumes high frequencies. A more complete
study of this specific system and a discussion regarding the physical
implications of this resonance are given in~\cite{ido}.

In this section we demonstrated that the effective Hamiltonian
$\hat{G}$
can be used to obtain some physical properties of systems that are
more complicated than those presented in Secs.~\ref{example1} and \ref{spin}.
In particular, the resonances of a periodic time dependent system were 
found to be given
by the resonances of the corresponding time independent effective Hamiltonian.
Resonances for oscillating barriers were computed numerically in~\cite{bagwell92,wagner97}.
The calculation of the present section demonstrates their physical origin.

\section{Summary and discussion}
\label{discussion}

In this paper we investigated classical and quantum motion in high
frequency fields. The classical motion can be treated by separation
of time scales. In Sec.~\ref{classical} this motion is separated into a slow part and
a fast part, which consists of rapid oscillations around the slow part.
The fast part and the resulting equation for the slow
motion are solved perturbatively to order $\omega^{-4}$. This
perturbation series is a generalization of the calculation
presented in {\em Mechanics} by Landau and Lifshitz~\cite{LL1}. We also demonstrated
that this perturbation theory is equivalent (to order $\omega^{-4}$) to the standard
mathematical method of multiple time scales
analysis~\cite{averaging}, which is more complicated.
The resulting equation for the slow motion is found to result from
a time independent Hamiltonian. 

Following a review of Floquet theory in Sec.~\ref{theory}
the corresponding  effective quantum Hamiltonian is computed explicitly, using a high frequency
perturbation theory up to order $\omega^{-4}$, in Sec.~\ref{highfreq}.
The resulting Hamiltonian (\ref{finalg}) is rather simple. Its classical limit
is the classical effective Hamiltonian~(\ref{classicalheff}).
This effective Hamiltonian is therefore a generalization of
the classical results of Kapitza~\cite{kapitza} and Landau and Lifshitz~\cite{LL1} to quantum mechanics.
In the classical limit the unitary gauge transformation $e^{-i \hat{F}}$ reduces
to a canonical transformation generated by the classical limit of $-\hbar \hat{F}$
(see (\ref{classicalg}) and \cite{GR}).
The Hamiltonian (\ref{classicalheff}) for the slow variables was
obtained in a natural way in Sec.~\ref{classical}.
The freedom in the choice of gauge was used to choose $\hat{F}$ so that in the classical
limit the effective quantum Hamiltonian (\ref{finalg}) reduces to (\ref{classicalheff}).
Consequently the classical limit of $-\hbar \hat{F}$ is the generating function of
the canonical transformation
from the original time dependent Hamiltonian to the time independent Hamiltonian (\ref{classicalheff}).
This limit explains the fact that the classical dynamics of the slow coordinate $X$ is
generated by a Hamiltonian. Using the freedom in the choice of gauge one can generate Hamiltonians
that differ from (\ref{classicalheff}) and (\ref{finalg}) but are related to them by canonical
and gauge transformations. 
The present work extends \cite{GR} to general driving potentials and is not
restricted to the driving (6a) of \cite{GR}.
The perturbation theory which was developed can, in principle, be used
to compute it to any given order in $1/\omega$.
This is a significant extension beyond \cite{GR} in the spirit of
separation of time scales~\cite{averaging} that enables a systematic expansion in powers
of $\omega^{-1}$. For this, the requirement (23) of \cite{GR} is avoided
and the expansion can be performed for any driving potential $V_1$ and is
not restricted to driving forces that are uniform in space.
It should be emphasized that this perturbation theory is an expansion
in $1/\omega$ and {\em not} in powers of the time dependent potential.
The potential $V_1$ does not have to be small in order to obtain
a good approximation of the original system.

Several examples were discussed. The driven harmonic oscillator and the
spin in a rotating magnetic field are simple, exactly solvable examples,
that were used as demonstrations for the method. For another system, the oscillating
Gaussian, we showed numerically that its lowest resonance is
given by the resonance of the corresponding effective Hamiltonian~(\ref{finalg}) of Sec.~\ref{gaussian}.
Thus, for time dependent traps, such as the atomic billiards discussed
earlier, the time independent effective Hamiltonian can be used
to compute the resonances and the lifetimes of particles in these traps.

While the examples presented in this work indicate that this effective
potential is a meaningful concept and is also useful for calculations, there are 
points that require further research. The convergence properties of the $1/\omega$
expansions for $\hat{F}$ and $\hat{G}$ are not clear. There may be situations
in which the perturbation theory fails to converge or where
smaller than any power corrections are of physical importance.
Consider for example a system similar to the oscillating Gaussian
where $V_1$ vanishes outside some finite domain. In this case one may expect
to find scattering states as the eigenstates of $\hat{G}$. In the
limit of high energies such a state may be roughly approximated by 
the plane wave $e^{ipx/\hbar}$ (with large $p$). 
The leading momentum contributions to $\hat{F}$ are of the form
\begin{eqnarray}
 \frac{1}{\hbar \omega} \int^\tau \left[ V_1 \right] \! -\! \frac{1}{\hbar m \omega^2} \int^{(2)\tau} \left[V_1^\prime \right] \hat{p}\! + \! \frac{1}{\hbar m^2 \omega^3} \int^{(3)\tau} \left[ V_1^{\prime \prime} \right] \hat{p}^2 \nonumber \\  - \frac{1}{\hbar m^3 \omega^4} \int^{(4)\tau} \left[ V_1^{(3)} \right] \hat{p}^3 \nonumber
\end{eqnarray}
as one can see from (\ref{finalf}) with (\ref{f3}) and (\ref{f4}).
While the general structure of the series for $\hat{F}$ is unknown, it seems
that for states with momentum which scales as $\omega^\alpha$ with $\alpha>1$
the series do not converge.

Even taking into account all the questions concerning the validity and convergence
of the perturbation expansions presented in this work we find the effective
potential to be useful for the following reasons.
The perturbation theory leads to a time independent effective Hamiltonian.
Physicists, who are used to work with time independent systems,
have developed an intuition for such systems, and thus
the effective Hamiltonian may give physical insight that is absent
when examining the corresponding time dependent problem.
In addition, all the calculations used to obtain the effective
Hamiltonian are straight forward. There are no differential equations
or any complicated iterative schemes that may appear in more
sophisticated perturbation theories. For comparison see~\cite{scherer95}.
Finally, one may use all the well developed techniques for time independent 
quantum systems to compute the eigenvalues of $\hat{G}$.
In particular, one can use time independent perturbation theory
in the case where the eigenvalues and eigenstates of $\hat{G}_0$ are known.

\begin{figure}[htb]
\includegraphics[width=7cm,height=5cm]{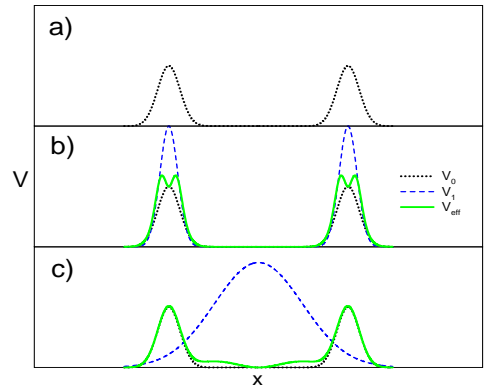}
\caption{Qualitative effect of driving on trapping times: a) the average potential $V_0$, b) the position dependence of the driving potential $V_1$ (dashed line) results in the effective potential $V_{eff}$ (solid line), c) as in b) with driving mainly inside the trap.  The units are arbitrary.\label{quality}}
\end{figure}

The effective Hamiltonian (\ref{finalg}) can be useful to predict
the qualitative behavior without complicated numerical calculations.
Assume first that the frequency is sufficiently high so that
only terms of order up to $1/\omega^2$ should be included and denote
$V_2 (x) \equiv \frac{1}{2m\omega^2} \overline{\left( \int^\tau V_1^\prime\right)^2}$.
Let $V_0$ take the form depicted in Fig.~\ref{quality}a.
It exhibits resonances. A natural question is whether the line width
increases or decreases as a result of the time dependent potential.
It is clear that for a situation of Fig.~\ref{quality}b 
the line width decreases, since the particle has to tunnel through effective
barriers which are higher than those of $V_0$
because the effective potential is $V_{eff}=V_0+V_2$,
while for Fig.~\ref{quality}c the line width (typically) increases since the energy
of the resonances is shifted upwards by the time dependent perturbation.
Numerical calculations of the type presented in Sec.~\ref{gaussian} and in \cite{bagwell92}
and \cite{wagner97} should confirm these results. If the terms of order $\omega^{-4}$
are important more subtle considerations concerning the kinetic energy $\left< \hat{p}^2 \right>/2m$
and the semiclassical limit are required.

Since there are some physical situations where one expects that the 
perturbation theory fails it is of importance to find ways to improve it.
The goal is to be able to describe terms that are smaller than any power
in the perturbations. This might be possible using
super-convergent perturbation theory that was recently
applied to time dependent quantum systems~\cite{super}.
In this perturbation theory the small parameter is the size of the
time dependent potential. It is of interest to modify it to
a perturbation theory in $1/\omega$. Note that while this
super-convergent perturbation schemes have superior convergence
properties, they may turn out to be complicated for explicit
calculations.
Thus, one may lose the main advantage of the perturbation theory
presented in this work, its simplicity.
In addition, it is of interest to generalize the results of this
work to systems of higher dimension. Such a generalization should be straight forward.

In conclusion, we have investigated the dynamics of high frequency
driven classical and quantum systems. High frequency perturbation theory
was used to obtain an effective time independent Hamiltonian for the slow part of the 
classical and quantum motion. For quantum systems, the spectrum of this Hamiltonian is the quasienergy spectrum of the
time dependent system. This effective Hamiltonian is computed in
a high frequency systematic perturbation theory.
It is demonstrated that the effective Hamiltonian gives the exact
quasienergies and quasienergy states of some simple examples as 
well as the lowest resonance (including the lifetime) for a time
dependent atom trap. 

\begin{acknowledgments}
It is our great pleasure to thank Michael V. Berry, Nir Davidson, Nimrod Moiseyev, and Vered Rom-Kedar
for stimulating and inspiring discussions and a referee of Phys. Rev. A. for bringing Refs. \cite{GR},
\cite{bagwell92} and \cite{wagner97} to our attention.
We thank Nimrod Moiseyev also for the involvement in many of the fine details
of this work.
 This research was supported in part by the US-Israel Binational
Science
Foundation (BSF), by the Fund for Promotion of Research at the Technion,
and by the Minerva Center of Nonlinear Physics of Complex Systems.
\end{acknowledgments}

\begin{appendix}
\section{Multiple time scales analysis}
\label{multiple}

The derivation given in Sec.~\ref{classical} was, in some sense, not explicitly consistent. For example the slow motion was not solved by expanding its
coordinates
in orders of $1/\omega$.
Terms of order $1/\omega^n$ included all contributions up to that order,
and also some contributions of higher order.
 It is of interest to show that the same result
can be obtained using a standard method, namely the method of multiple time scales analysis~\cite{averaging}. 
In this appendix we show how to derive the equations of motion of
the slow dynamics using multiple time scales analysis. We will present
only the first few orders in $1/\omega$ since this method turns out to
be more complicated than the method used in Sec.~\ref{classical}.
While the two methods differ in details they are equivalent and lead to the same results, when consistently expanded order by order in $1/\omega$.

In order to use multiple time scales analysis it is convenient to 
transform equation~(\ref{newton}) to a standard form~\cite{averaging}.
This can be done by defining $\tau \equiv \omega t$, $\epsilon \equiv \frac{1}{\omega}$ and $y \equiv \frac{dx}{dt}=\omega \frac{dx}{d\tau}$. The 
first order equations of
motion are then given by
\begin{equation}
\label{cform}
 \frac{d}{d \tau}\vec{z} = \epsilon \vec{f} (\tau, \vec{z}) 
\end{equation}
where
\begin{equation}
\vec{z}= \left( \begin{array}{c} x \\ y \end{array} \right)
\end{equation}
and
\begin{equation}
\label{epsexp}
\vec{f}= \left( \begin{array}{c} y \\ -\frac{1}{m} \left( V_0^\prime (x) + V_1^\prime (x,\tau) \right) \end{array} \right).
\end{equation}
Then one introduces the following expansion of the solution
\begin{equation}
 \vec{z} = \sum_{n=0}^{\infty} \epsilon^n \vec{z}_n (\tau,t),
\end{equation}
where $t=\epsilon \tau$ and $\tau$ are treated as independent variables. 
Using this expansion together with 
\begin{equation}
\frac{d}{d \tau}=\frac{\partial}{\partial \tau}+\epsilon \frac{\partial}{\partial t}  
\end{equation}
results in
\begin{equation}
\frac{\partial \vec{z}_0}{\partial \tau} + \epsilon \frac{\partial \vec{z}_0}{\partial t}+ \epsilon \frac{\partial \vec{z}_1}{\partial \tau} + \cdots = \epsilon \vec{f} (\tau, \vec{z}_0 + \epsilon \vec{z}_1 + \cdots).
\end{equation}
The solution is obtained by expanding $\vec{f}$, matching powers of $\epsilon$,
and solving for $\vec{z}$ order by order. This is the standard multiple 
time scales analysis, see~\cite{averaging} for a detailed description of
the method. (Note that in~\cite{averaging} the role of $t$ and $\tau$
is opposite to the one in the present work). 
We proceed to solve the first few orders in $\epsilon$.

At order $\epsilon^0$ the leading order of (\ref{epsexp}) results in
\begin{eqnarray}
\frac{\partial x_0}{\partial \tau} & = & 0 \nonumber \\
\frac{\partial y_0}{\partial \tau} & = & 0.
\end{eqnarray}
Therefore $x_0$ and $y_0$ can be any functions of the slow time $t$
\begin{eqnarray}
\label{eps0sol}
x_0 & = & \bar{x}_0 (t) \nonumber \\
y_0 & = & \bar{y}_0 (t).
\end{eqnarray} 
Note that at the leading order the solution is found to depend only on the slow time scale, as expected.
We will denote the $\tau$ independent slow part of the solution by
$\bar{x}_i$ and $\bar{y}_i$ at any order.
Additional conditions (\ref{secular1}) on $\bar{x}_0$ and $\bar{y}_0$ will be obtained from the
requirement that the solution is not secular at the next order.

At order $\epsilon$ (\ref{epsexp}) results in
\begin{eqnarray}
\label{eps1}
\frac{\partial x_1}{\partial \tau} & = & - \frac{\partial \bar{x}_0}{\partial t} + \bar{y}_0 (t) \nonumber \\
\frac{\partial y_1}{\partial \tau} & = & - \frac{\partial \bar{y}_0}{\partial t} -\frac{1}{m} V_0^\prime (\bar{x}_0) -\frac{1}{m} V_1^\prime (\bar{x}_0,\tau),
\end{eqnarray}
where we have used (\ref{eps0sol}).
We are interested in solutions of (\ref{eps1}) that are not secular, namely, that do
not grow with the fast time scale $\tau$. The RHS of equation~(\ref{eps1})
can be decomposed into periodic functions of $\tau$ (with vanishing average) and $\tau$ independent terms.
Any $\tau$ independent term will result in a secular contribution.
Therefore, to avoid such terms we demand
\begin{eqnarray}
\label{secular1}
 \frac{ \partial \bar{x}_0}{\partial t} & = & \bar{y}_0 (t) \nonumber \\
 \frac{ \partial \bar{y}_0}{\partial t} & = & - \frac{1}{m} V_0^\prime (\bar{x}_0 (t)).
\end{eqnarray}
This is the leading order of the equation that governs the slow time scale.
The non secular equation of order $\epsilon$ is now
\begin{eqnarray}
\frac{\partial x_1}{\partial \tau} & = & 0 \nonumber \\
\frac{\partial y_1}{\partial \tau} & = & -\frac{1}{m} V_1^\prime (\bar{x}_0,\tau)
\end{eqnarray} 
and its solution is
\begin{eqnarray}
x_1 (\tau,t) & = & \bar{x}_1 (t) \nonumber \\
y_1 (\tau,t) & = & -\frac{1}{m}\int^\tau \left[ V_1^\prime (\bar{x}_0 (t) ,\tau)\right] + \bar{y}_1 (t).
\end{eqnarray} 

This process can be repeated order after order. At each order all the terms
with the same power of
 $\epsilon$ are gathered. The resulting equation may
be secular and therefore an additional condition on the slow part of the solution
is enforced. Then, one can solve for $x$ and $y$ at that order. 
The calculation is rather tedious and will not be presented here. We only
give the secularity conditions that result when the next two orders are computed:
\begin{eqnarray}
\label{secular2}
 \frac{ \partial \bar{x}_1}{\partial t} & = & \bar{y}_1 (t) \nonumber \\
 \frac{ \partial \bar{y}_1}{\partial t} & = & - \frac{1}{m}\bar{x}_1 (t) V_0^{\prime \prime} (\bar{x}_0 (t)),
\end{eqnarray}
and
\begin{eqnarray}
\label{secular3}
 \frac{ \partial \bar{x}_2}{\partial t} & = & \bar{y}_2 (t) \\
 \frac{ \partial \bar{y}_2}{\partial t} & = & \frac{1}{m^2} \overline{V_1^{\prime \prime} (\bar{x}_0(t),\tau) \int^{(2)\tau} \left[ V_1^\prime (\bar{x}_0(t),\tau)\right]} \nonumber \\ & - & \frac{1}{m} \bar{x}_2 (t) V_0^{\prime \prime} (\bar{x}_0 (t)) - \frac{\bar{x}_1^2 (t)}{2m} V_0^{(3)} (\bar{x_0}(t)) . \nonumber
\end{eqnarray}

Equations (\ref{secular1}), (\ref{secular2}) and (\ref{secular3}) 
are the first orders of an
expansion of the following system of equations
\begin{eqnarray}
\label{summed}
\frac{\partial \bar{X}}{\partial t} & = & \bar{Y} (t) \nonumber \\
\frac{\partial \bar{Y}}{\partial t} & = & - \frac{1}{m} V_0^\prime (\bar{X} (t)) \nonumber \\ & + & \frac{\epsilon^2}{m^2}  \overline{V_1^{\prime \prime} (\bar{X}(t),\tau) \int^{(2)\tau} \left[ V_1^\prime (\bar{X}(t),\tau)\right]}
\end{eqnarray}
in powers of $\epsilon$ where
\begin{eqnarray}
\label{slowexpansion}
\bar{X}(t) & \equiv & \bar{x}_0 (t)+ \epsilon \bar{x}_1 (t)+ \epsilon^2 \bar{x}_2(t) + \cdots \nonumber \\
\bar{Y}(t) & \equiv & \bar{y}_0(t) + \epsilon \bar{y}_1(t) + \epsilon^2 \bar{y}_2(t) + \cdots
\end{eqnarray}
Equations (\ref{summed}) are accurate to order $\epsilon^2$ and are identical to the slow equation~(\ref{slow2}). 
We have used the multiple time scales analysis up to order $\epsilon^4$ and found that the resulting slow equations analogous to (\ref{summed})
are equivalent to equation~(\ref{slow4}) in Sec.~\ref{classical}, with the
identification $\bar{X}=X$ and $\bar{Y}=\dot{X}$.

The method presented in Sec.~\ref{classical} and multiple time scales analysis used in this appendix have a similar structure. Both separate the
motion into $\tau$ dependent and independent parts and both lead to equations
for the slow motion that have to be satisfied to avoid solutions that grow
with $\tau$. The main difference between the methods is that when using
multiple time scales the slow coordinate is expanded in powers of $\epsilon=1/\omega$.  
This leads to a large number of terms that result of the fact that the functions
in the slow equation are evaluated at $\bar{x}_0(t)$ rather than at $\bar{X}(t)$.
Consequently the derivation in Sec.~\ref{classical} is much simpler
than the one presented in this appendix.
This is also the reason that here we did not present explicitly the calculation using
multiple time scales analysis up to order $\epsilon^4$.

\section{Some useful operator relations}
\label{operator}

In this appendix some known relations (see for example~\cite{messiah}) involving operators
and exponentials of other operators will be presented for completeness.
Let us define
\begin{equation}
\hat{C} (x) = e^{x \hat{A}} \hat{B} e^{-x \hat{A}}.
\end{equation}
It is clear that $\hat{C} (0) = \hat{B}$. Differentiation  
shows that
\begin{equation}
\frac{d \hat{C}}{d x} = \left[ \hat{A}, \hat{C}\right].
\end{equation}
Therefore $\hat{C} (x)$ has the following expansion in powers of
$x$
\begin{eqnarray}
\label{xepansion}
\hat{C} & = & \hat{B}+ x \left[\hat{A},\hat{B} \right] + \frac{1}{2} x^2 \left[ \hat{A}, \left[\hat{A},\hat{B} \right]\right] \nonumber \\ & + & \frac{1}{6} x^3 \left[ \hat{A},\left[ \hat{A}, \left[\hat{A},\hat{B} \right]\right]\right] + \cdots 
\end{eqnarray}
Substitution of $x=1$, $\hat{A}=i \hat{F}$ and $\hat{B}=\hat{H}$
results in~(\ref{eifhemif}).

One can use the expansion~(\ref{eifhemif}) to find an expansion
of $\left(\frac{\partial e^{i\hat{F}}}{\partial \tau} \right) e^{-i \hat{F}}$.
First, since $e^{i \hat{F}} e^{-i \hat{F}}= 1$,
\begin{equation}
\label{timeder}
 \left(\frac{\partial e^{i\hat{F}}}{\partial \tau} \right) e^{-i \hat{F}} =  -e^{i\hat{F}} \left( \frac{\partial e^{-i\hat{F}}}{\partial \tau} \right)=  -e^{i\hat{F}}  \hat{\frac{\partial }{\partial \tau}} e^{-i\hat{F}} + \hat{\frac{\partial }{\partial \tau}},
\end{equation} 
where in the last expression the derivative $\hat{\frac{\partial }{\partial \tau}}$ in front of $e^{-i \hat{F}}$
operates not only on the exponential but also on any operator or function
that appears on its right. Using (\ref{xepansion}) with $\hat{B}=\hat{\frac{\partial}{\partial \tau}}$ and $\hat{A}=i \hat{F}$
in (\ref{timeder}) yields
\begin{eqnarray}
\label{timeder2}
\left(\frac{\partial e^{i\hat{F}}}{\partial \tau} \right) e^{-i \hat{F}}&  = &- i \left[ \hat{F}, \hat{\frac{\partial }{\partial \tau}}\right] + \frac{1}{2} \left[ \hat{F}, \left[ \hat{F}, \hat{\frac{\partial }{\partial \tau}}\right]\right] \nonumber \\ & + & \frac{i}{3!} \left[ \hat{F},\left[ \hat{F}, \left[ \hat{F}, \hat{\frac{\partial }{\partial \tau}}\right]\right]\right] + \cdots 
\end{eqnarray}
Substitution of
\begin{equation}
\label{timecomm}
\left[ \hat{F}, \hat{\frac{\partial }{\partial \tau}}\right]= -\frac{\partial \hat{F}}{\partial \tau}
\end{equation}
in~(\ref{timeder2})
results in~(\ref{eifemif}).

If the operators $\hat{A}$ and $\hat{B}$ commute with
their commutator $\left[ \hat{A},\hat{B} \right]$, then~\cite{messiah}
\begin{equation}
\label{commute}
 e^{\hat{A}+\hat{B}} = e^{-\frac{1}{2}\left[ \hat{A},\hat{B} \right] } e^{\hat{A}} e^{\hat{B}}.
\end{equation}
This is a special case of the Campbell-Baker-Hausdorff formula~\cite{englman63}.

\section{Some high orders of the effective Hamiltonian}
\label{highorders}

 In this appendix we will outline the steps that should be
taken in order to compute $\hat{G}$ to order $\omega^{-4}$. 
The term of order $\omega^{-3}$ found from (\ref{effect2})-(\ref{eifemif}) is
\begin{widetext}
\begin{eqnarray}
\label{originalg3}
\hat{G}_3 & = & i \left[ \hat{F}_3, \hat{H} \right] - \frac{1}{2}  \left[\hat{F}_2,\left[ \hat{F}_1, \hat{H} \right] \right] - \frac{1}{2}  \left[\hat{F}_1,\left[ \hat{F}_2, \hat{H} \right] \right] - \frac{i}{6} \left[ \hat{F}_1,  \left[\hat{F}_1,\left[ \hat{F}_1, \hat{H} \right] \right] \right]  
-  \hbar \frac{\partial \hat{F}_4}{\partial \tau} -\frac{i\hbar}{2} \left[ \hat{F}_1, \frac{\partial \hat{F}_3}{\partial \tau} \right]  \nonumber \\ & - &\frac{i\hbar}{2} \left[ \hat{F}_2, \frac{\partial \hat{F}_2}{\partial \tau} \right]  \nonumber - \frac{i\hbar}{2} \left[ \hat{F}_3, \frac{\partial \hat{F}_1}{\partial \tau} \right]
+ \frac{\hbar}{6} \left[ \hat{F}_1,  \left[ \hat{F}_1, \frac{\partial \hat{F}_2}{\partial \tau} \right] \right] 
+  \frac{\hbar}{6} \left[ \hat{F}_1,  \left[ \hat{F}_2, \frac{\partial \hat{F}_1}{\partial \tau} \right] \right] \nonumber \\ & + &  \frac{\hbar}{6} \left[ \hat{F}_2,  \left[ \hat{F}_1, \frac{\partial \hat{F}_1}{\partial \tau} \right] \right] + \frac{i \hbar}{24} \left[ \hat{F}_1 ,\left[ \hat{F}_1,  \left[ \hat{F}_1, \frac{\partial \hat{F}_1}{\partial \tau} \right] \right] \right]. 
\end{eqnarray}
\end{widetext}
 It can be further simplified by taking into consideration that some
of the commutators vanish, for instance,
$ \left[ \hat{F}_1,  \left[ \hat{F}_2, \frac{\partial \hat{F}_1}{\partial \tau} \right] \right]=0$ since there is only one derivative (with respect to $x$) and two commutations.
Equations (\ref{df2dt}) and (\ref{df3dt}) for $\hat{F}_2$ and $\hat{F}_3$ can be used to eliminate the commutation relations that involve $\hat{H}$.
After some manipulations~(\ref{originalg3}) reduces to
\begin{equation}
\label{nextg3}
\hat{G}_3 = i \left[ \hat{F}_3 , \hat{G}_0\right] - \hbar \frac{\partial \hat{F}_4}{\partial \tau} + \frac{i \hbar}{2} \left[ \hat{F}_3, \frac{\partial \hat{F}_1}{\partial \tau} \right].
\end{equation}
$\hat{F}_4$ will be chosen so that the periodic terms (with vanishing average)
in (\ref{nextg3}) are eliminated. To examine the contributions to $\hat{G}_3$
which remain, one can just average (\ref{nextg3}) over a period.
This leads to
\begin{eqnarray}
\label{nng3}
\hat{G}_3 & = & i\overline{ \left[ \hat{F}_3 , \hat{G}_0\right]} +\frac{i \hbar}{2} \overline{\left[ \hat{F}_3, \frac{\partial \hat{F}_1}{\partial \tau} \right]} \nonumber \\
 & = & i \left[ \hat{f}_3, \frac{\hat{p}^2}{2m} + \hat{V}_0 \right] + \frac{i \hbar}{2 m^2} \left\{ 2 \overline{\int^\tau \left[ V_1^\prime\right] \int^{(2)\tau} \left[V_1^{\prime \prime} \right]} \frac{\partial}{\partial x} \right. \nonumber \\ & + & \left. \overline{\int^\tau \left[ V_1^\prime\right] \int^{(2)\tau} \left[V_1^{(3)} \right]} \right\}.
\end{eqnarray}
The freedom to choose $\hat{f}_3$ can be used to cause $\hat{G}_3$ to vanish.
We choose $\hat{f}_3$ to be a function of $x$ so that
\begin{equation}
\label{f3choice}
 f_3^\prime (x) = - \frac{1}{m \hbar} \overline{\int^\tau \left[ V_1^\prime\right] \int^{(2)\tau} \left[V_1^{\prime \prime} \right] }.
\end{equation}
Using (\ref{f3choice}) in (\ref{nng3}) indeed results in
\begin{equation}
\label{g3}
 \hat{G}_3=0.
\end{equation}
Thus we will use this choice of gauge. With this choice of $\hat{f}_3$ (and therefore of $\hat{F}_3$), we require that $\hat{F}_4$ will satisfy
\begin{equation}
\label{df4dt}
\frac{\partial \hat{F}_4}{\partial \tau} = \frac{i}{\hbar} \left[ \hat{F}_3, \hat{G}_0 \right] + \frac{i}{2} \left[ \hat{F}_3,\frac{\partial \hat{F}_1}{\partial \tau} \right]. 
\end{equation}
Equation (\ref{df4dt}) can be integrated in order to compute $\hat{F}_4$.
Using (\ref{g0}), (\ref{f3}) and (\ref{f3choice}) leads to
\begin{widetext}
\begin{eqnarray}
\label{f4}
 \hat{F}_4 & = & - \frac{\hbar^2 i}{m^3} \int^{(4)\tau} \left[ V_1^{(3)}\right]\frac{\partial^3}{\partial x^3} - \frac{3\hbar^2 i}{2m^3} \int^{(4)\tau} \left[ V_1^{(4)}\right]\frac{\partial^2}{\partial x^2} - \frac{3\hbar^2 i}{4m^3} \int^{(4)\tau} \left[ V_1^{(5)}\right]\frac{\partial}{\partial x} - \frac{\hbar^2 i}{8m^3} \int^{(4)\tau} \left[ V_1^{(6)}\right] \\
& - & \frac{i}{2 m^2} \left( \int^\tau \left[ {\cal P}_3^\prime \right] + \int^\tau \left[ {\cal P}_2^\prime \right] + 2  \int^\tau \left[ {\cal P}_3 \right] \frac{\partial}{\partial x}   + 2  \int^\tau \left[ {\cal P}_2 \right] \frac{\partial}{\partial x} \right) 
 +  \frac{i}{4 m^2} \left( \int^{(2)\tau} \left[ {\cal P}_1^{\prime \prime} \right] +2  \int^{(2)\tau} \left[ {\cal P}_1^{\prime} \right] \frac{\partial}{\partial x} \right) + \hat{f}_4 (\hat{x},\hat{p}), \nonumber
\end{eqnarray}
where $\hat{f}_4$ is a hermitian time independent operator (that depends only on
$\hat{x}$ and $\hat{p}$) 
that will be determined at the next stage while
\begin{eqnarray}
\label{defp23}
{\cal P}_2 (x,\tau) & \equiv & V_1^\prime \int^{(3)\tau} \left[V_1^{\prime \prime} \right] - \overline{V_1^\prime \int^{(3)\tau} \left[ V_1^{\prime \prime}\right]} \nonumber \\
{\cal P}_3 (x,\tau) & \equiv & V_0^{\prime \prime} \int^{(3)\tau} \left[ V_1^\prime \right] + 3 V_0^\prime \int^{(3)\tau} \left[ V_1^{\prime \prime} \right].
\end{eqnarray}
The operator $\hat{f}_4$ plays here a role similar to the one of $\hat{f}_3$.
Note that $\overline{{\cal P}_1}= \overline{{\cal P}_2}=\overline{{\cal P}_3}=0$.

We turn to compute the next order of $\hat{G}$, order $\omega^{-4}$.
This is the last order that will be considered explicitly in this work. The terms of order
$\omega^{-4}$ in (\ref{effect2})-(\ref{eifemif}) are given by
\begin{eqnarray}
\label{originalg4}
\hat{G}_4 & = & i \left[ \hat{F}_4, \hat{H} \right] -\frac{1}{2} \hat{{\cal L}_1}   -\frac{i}{6}\hat{{\cal L}_2}   + \frac{1}{24} \left[ \hat{F}_1,\left[ \hat{F}_1,\left[ \hat{F}_1 , \left[ \hat{F}_1, \hat{H} \right] \right]\right]\right]
- \hbar \frac{\partial \hat{F}_5}{ \partial \tau} 
  -   \frac{i \hbar}{2} \hat{{\cal M}_1} 
+ \frac{\hbar}{6}\hat{{\cal M}_2}  + \frac{i \hbar}{24}\hat{{\cal M}_3}  \nonumber \\ & - & \frac{\hbar}{120} \left[ \hat{F}_1,\left[ \hat{F}_1 , \left[ \hat{F}_1 , \left[ \hat{F}_1,\frac{\partial \hat{F}_1}{\partial \tau } \right]\right] \right] \right].
\end{eqnarray}
where
\begin{eqnarray}
 \hat{{\cal L}_1} &  = & \left[ \hat{F}_1 , \left[ \hat{F}_3, \hat{H} \right] \right] +\left[ \hat{F}_2 , \left[ \hat{F}_2, \hat{H} \right] \right] +\left[ \hat{F}_3 , \left[ \hat{F}_1, \hat{H} \right] \right], \\
 \hat{{\cal L}_2} & = &\left[ \hat{F}_1,\left[ \hat{F}_1 , \left[ \hat{F}_2, \hat{H} \right] \right]\right] + \left[ \hat{F}_1,\left[ \hat{F}_2 , \left[ \hat{F}_1, \hat{H} \right] \right]\right] + \left[ \hat{F}_2,\left[ \hat{F}_1 , \left[ \hat{F}_1, \hat{H} \right] \right]\right], \\
\hat{{\cal M}_1} & = &\left[ \hat{F}_4,\frac{\partial \hat{F}_1}{ \partial \tau} \right] +\left[ \hat{F}_3,\frac{\partial \hat{F}_2}{ \partial \tau} \right]  +\left[ \hat{F}_2,\frac{\partial \hat{F}_3}{ \partial \tau} \right] +\left[ \hat{F}_1,\frac{\partial \hat{F}_4}{ \partial \tau} \right], \\
\hat{{\cal M}_2} & = & \left[ \hat{F}_1 , \left[ \hat{F}_1,\frac{\partial \hat{F}_3}{ \partial \tau} \right]\right] +  \left[ \hat{F}_1 , \left[ \hat{F}_3,\frac{\partial \hat{F}_1}{ \partial \tau} \right]\right] + \left[ \hat{F}_3 , \left[ \hat{F}_1,\frac{\partial \hat{F}_1}{ \partial \tau} \right]\right] 
+ \left[ \hat{F}_1 , \left[ \hat{F}_2,\frac{\partial \hat{F}_2}{ \partial \tau} \right]\right] \nonumber \\ & + & \left[ \hat{F}_2 , \left[ \hat{F}_1,\frac{\partial \hat{F}_2}{ \partial \tau} \right]\right] + \left[ \hat{F}_2 , \left[ \hat{F}_2,\frac{\partial \hat{F}_1}{ \partial \tau} \right]\right],
\end{eqnarray}
and
\begin{equation}
\hat{{\cal M}_3}  =  \left[ \hat{F}_1 , \left[ \hat{F}_1 , \left[ \hat{F}_1,\frac{\partial \hat{F}_2}{ \partial \tau} \right]\right] \right] + \left[ \hat{F}_1 , \left[ \hat{F}_1 , \left[ \hat{F}_2,\frac{\partial \hat{F}_1}{ \partial \tau} \right]\right] \right] 
 + \left[ \hat{F}_1 , \left[ \hat{F}_2 , \left[ \hat{F}_1,\frac{\partial \hat{F}_1}{ \partial \tau} \right]\right] \right] + \left[ \hat{F}_2 , \left[ \hat{F}_1 , \left[ \hat{F}_1,\frac{\partial \hat{F}_1}{ \partial \tau} \right]\right] \right].
\end{equation}
Many commutation relations vanish since $\hat{F}_2$ contains one derivative
(see (\ref{f2})) and $\hat{F}_1$ contains no derivatives. Repeated
application of the commutation relations results in the vanishing
of such relations.
The commutation relations of $\hat{H}$ (except $\left[ \hat{F}_4,\hat{H}\right]$)
can be eliminated using (\ref{f1}), (\ref{df2dt}), (\ref{df3dt}) and (\ref{df4dt}).
After a tedious calculation, equation (\ref{originalg4}) can be simplified to
\begin{equation}
\label{nextg4}
\hat{G}_4 =  i \left[ \hat{F}_4, \hat{G}_0 \right] - \frac{\hbar}{4} \left[\hat{F}_2,\overline{\left[ \hat{F}_2, \frac{\partial \hat{F}_1}{\partial \tau} \right]} \right] - \hbar \frac{\partial \hat{F}_5}{\partial \tau} + \frac{i \hbar}{2} \left[ \hat{F}_4, \frac{\partial \hat{F}_1}{\partial \tau} \right] 
 -  \frac{\hbar}{12} \left[ \hat{F}_1, \left[ \hat{F}_3, \frac{\partial \hat{F}_1}{\partial \tau} \right] \right] - \frac{\hbar}{12} \left[ \hat{F}_2, \left[ \hat{F}_2, \frac{\partial \hat{F}_1}{\partial \tau} \right] \right].
\end{equation}
Our goal is to obtain an explicit expression for $\hat{G}_4$ as was done
for lower orders. Note that we do not have to compute
$\hat{F}_5$ explicitly for this purpose since $\hat{F}_5$ is chosen in such
a way that it will cancel the time dependent part of (\ref{nextg4})
and it is required only for the calculation of higher order
terms of $\hat{G}$.
The terms that are not canceled are just the time independent terms
on the RHS of (\ref{nextg4}). They are obtained simply
by averaging (\ref{nextg4}) over a period resulting in
\begin{equation}
\label{avgg4-}
\hat{G}_4 = i \left[ \hat{f}_4, \hat{G}_0 \right] + \frac{i \hbar}{2} \overline{\left[ \hat{F}_4, \frac{\partial \hat{F}_1}{\partial \tau} \right] } -\frac{\hbar}{12} \overline{\left[ \hat{F}_1, \left[ \hat{F}_3, \frac{\partial \hat{F}_1}{\partial \tau} \right] \right]} -\frac{\hbar}{12} \overline{\left[ \hat{F}_2, \left[ \hat{F}_2, \frac{\partial \hat{F}_1}{\partial \tau} \right] \right]}.
\end{equation}
Substitution of the expressions for $\hat{F}_i$ and $\hat{G}_0$
leads after some straight forward but tedious calculations to
\begin{eqnarray}
\label{avgg4}
\hat{G}_4 & = & i \left[ \hat{f}_4 , \frac{\hat{p}^2}{2m} + V_0 \right] + \frac{\hbar^2}{2m^3} \left\{ 3 \overline{\int^{(2)\tau} \left[ V_1^\prime\right] \int^{(2)\tau}\left[V_1^{(3)}\right]}\frac{\partial^2}{\partial x^2} +3 \left( \overline{\int^{(2)\tau}\left[ V_1^{\prime \prime}\right]\int^{(2)\tau}\left[ V_1^{(3)}\right]}
\right. \right. \nonumber \\ & + & \left. \left. \overline{\int^{(2)\tau}\left[ V_1^{\prime}\right]\int^{(2)\tau}\left[ V_1^{(4)}\right]}\right)\frac{\partial}{\partial x}+ \overline{\left(\int^{(2)\tau} \left[ V_1^{(3)}\right]\right)^2} 
+ \frac{3}{2} \overline{\int^{(2)\tau}\left[ V_1^{\prime \prime}\right]\int^{(2)\tau}\left[ V_1^{(4)}\right]} \right. \nonumber \\ 
& + & \left. \frac{3}{4} \overline{\int^{(2)\tau}\left[ V_1^{\prime}\right]\int^{(2)\tau}\left[ V_1^{(5)}\right]}\right\} + \frac{1}{2m^2} \left\{ V_0^{\prime \prime} \overline{\left(\int^{(2)\tau} \left[ V_1^{\prime}\right]\right)^2 } + 3 V_0^\prime \overline{\int^{(2)\tau}\left[ V_1^{\prime}\right]\int^{(2)\tau}\left[ V_1^{\prime \prime}\right]} \right\} \nonumber \\
& + & \frac{1}{ 3 m^2} \left\{ \overline{V_1^{\prime \prime} \left( \int^{(2)\tau}\left[ V_1^\prime\right]\right)^2} +\overline{ V_1^{\prime } \int^{(2)\tau}\left[V_1^\prime \right] \int^{(2)\tau} \left[ V_1^{\prime \prime}\right]} -2 \overline{V_1^\prime \int^\tau \left[ V_1^\prime\right] \int^{(3)\tau} \left[V_1^{\prime \prime} \right]}\right\}.
\end{eqnarray}
Using integration by parts it is possible to show that
\begin{equation}
\label{simpl}
\overline{ V_1^{\prime } \int^{(2)\tau}\left[V_1^\prime \right] \int^{(2)\tau} \left[ V_1^{\prime \prime}\right]} -2 \overline{V_1^\prime \int^\tau \left[ V_1^\prime\right] \int^{(3)\tau} \left[V_1^{\prime \prime} \right]} = \frac{1}{2}\overline{V_1^{\prime \prime} \left( \int^{(2)\tau}\left[ V_1^\prime\right]\right)^2}. 
\end{equation}
This will simplify the last term
of (\ref{avgg4}). 
The operator $\hat{f}_4$ is not determined yet. We choose it of the form:
\begin{equation}
\label{f4small}
 \hat{f}_4 = \tilde{g}(x) \hat{p} + \hat{p} \tilde{g}(x)= 2 \frac{\hbar}{i} \tilde{g}(x) \frac{\partial}{\partial x} + \frac{\hbar}{i} \tilde{g}^\prime (x)
\end{equation}
where $\tilde{g}(x)$ is a function that should be determined.
In terms of the function $\tilde{g}(x)$ we find
\begin{eqnarray}
\label{g4withg}
\hat{G}_4 & = & 2 \hbar \tilde{g} V_0^\prime + \frac{\hbar^3}{2 m} \left( 4 \tilde{g}^\prime \frac{\partial^2}{\partial x^2} + 4 \tilde{g}^{\prime \prime} \frac{\partial}{\partial x} + \tilde{g}^{(3)} \right) + \frac{\hbar^2}{2 m^3} \left\{ 3 \overline{\int^{(2)\tau} \left[ V_1^\prime \right] \int^{(2)\tau} \left[ V_1^{(3)} \right] } \frac{\partial^2}{\partial x^2} \right. \nonumber \\
& + & \left. 3 \left( \overline{\int^{(2)\tau} \left[ V_1^{\prime \prime} \right] \int^{(2)\tau} \left[ V_1^{(3)} \right] }+\overline{\int^{(2)\tau} \left[ V_1^{\prime} \right] \int^{(2)\tau} \left[ V_1^{(4)} \right] } \right) \frac{\partial}{\partial x} + \overline{\left( \int^{(2)\tau} \left[ V_1^{(3)} \right]\right)^2} \right. \nonumber \\ 
& + & \left.  \frac{3}{2} \overline{\int^{(2)\tau} \left[ V_1^{\prime \prime} \right] \int^{(2)\tau} \left[ V_1^{(4)} \right] }+ \frac{3}{4} \overline{\int^{(2)\tau} \left[ V_1^{\prime} \right] \int^{(2)\tau} \left[ V_1^{(5)} \right] } \right\} + \frac{1}{2m^2} \overline{V_1^{\prime \prime} \left(\int^{(2)\tau} V_1^\prime \right)^2} \nonumber \\
& + & \frac{1}{2m^2} \left\{ V_0^{\prime \prime} \overline{ \left( \int^{(2)\tau} \left[ V_1^\prime \right] \right)^2 } + 3 V_0^\prime \overline{ \int^{(2)\tau} \left[ V_1^\prime \right] \int^{(2)\tau} \left[ V_1^{\prime \prime} \right]} \right\}.
\end{eqnarray}
We choose $\tilde{g}(x)$ so that in the classical limit the effective
Hamiltonian $\hat{G}$ reduces to its classical counterpart (\ref{classicalheff}).
It takes the form
\begin{equation}
\label{functiong}
\tilde{g}(x) = - \frac{3}{4 m^2 \hbar} \overline{\int^{(2)\tau} \left[ V_1^\prime\right]\int^{(2)\tau} \left[ V_1^{\prime \prime}\right]}.
\end{equation}
The resulting operator $\hat{G}_4$ is obtained by substituting (\ref{functiong}) in (\ref{g4withg}). It results
in the term of order $\omega^{-4}$ in the effective Hamiltonian (\ref{finalg}):
\begin{eqnarray}
\label{g4}
\hat{G}_4 & = & \frac{1}{2 m^2} V_0^{\prime \prime} \overline{\left( \int^{(2)\tau} \left[ V_1^\prime \right]\right)^2} + \frac{1}{2 m^2} \overline{V_1^{\prime \prime}\left( \int^{(2)\tau} \left[ V_1^\prime \right]\right)^2 } -  \frac{\hbar^2}{2 m^3} \left\{ 3 \overline{\left( \int^{(2)\tau} \left[ V_1^{\prime \prime} \right]\right)^2} \frac{\partial^2}{\partial x^2} \right. \nonumber \\ & + & \left. 6 \overline{\int^{(2)\tau} \left[ V_1^{\prime \prime}\right] \int^{(2)\tau} \left[ V_1^{(3)}\right]}\frac{\partial}{\partial x} + \frac{3}{2}\overline{\int^{(2)\tau} \left[ V_1^{\prime \prime}\right]\int^{(2)\tau} \left[ V_1^{(4)}\right]} + \frac{5}{4} \overline{\left( \int^{(2)\tau} \left[ V_1^{(3)} \right]\right)^2} \right\}.
\end{eqnarray}
This is the highest order of $\hat{G}$ that is computed here.
\end{widetext}
\end{appendix}

\typeout{References}


\begin{thebibliography}{99}

\bibitem{cornell02}
E. A. Cornell and C. E. Wieman,
{\em Rev. Mod. Phys.}, {\bf 74}, 875 (2002).

\bibitem{tannudjibook2}
C. Cohen-Tannoudji, J. Dupont-Roc and G. Grynberg,
{\em Atom-photon interactions}, (Wiely, New York, 1992).

\bibitem{faisal}
F. H. M. Faisal, {\em Theory of multiphoton processes},
(Plenum, New York, 1987).

\bibitem{raizen01}
V. Milner, J. L. Hanssen, W. C. Campbell and M. G. Raizen,
{\em Phys. Rev. Lett.}, {\bf 86}, 1514 (2001).

\bibitem{davidson01}
N. Friedman, A. Kaplan, D. Carasso and N. Davidson,
{\em Phys. Rev. Lett.}, {\bf 86}, 1518 (2001).

\bibitem{kapitza}
D. ter Haar (Ed.), {\em Collected papers of P. L. Kapitza}, (Pergamon Press,
Oxford, 1965). P. L. Kapitza, {\em Zh. Ek. Te. Fi.}, {\bf 21}, 588 (1951).

\bibitem{LL1}
L. D. Landau and E. M. Lifshitz, {\em Mechanics}, (Pergamon Press,
Oxford, 1976).

\bibitem{percival}
I. C. Percival and D. Richards, {\em Introduction to Dynamics},
(Cambridge University Press, London, 1982).

\bibitem{paul}
W. Paul, {\em Rev. Mov. Phys.}, {\bf 62}, 531 (1990).

\bibitem{glauber95}
G. Schrage, V. I. Man'ko, W. P. Schleich and R. J. Glauber,
{\em Quant. Semiclass. Opt.}, {\bf 7}, 307 (1995).

\bibitem{perelomov69}
A. M. Perelomov and V. S. Popov, {\em Sov. J. Theor. Math. Phys.}, {\bf 1}, 275 (1969).

\bibitem{perelomovbook}
A. M. Perelomov and Y. B. Zeldovich, {\em Quantum mechanics - selected topics},
 (World Scientific, Singapore, 1998).

\bibitem{GR}
T. P. Grozdanov and M. J. Rakovi\'c, {\em Phys. Rev. A},
{\bf 38}, 1739 (1988).

\bibitem{gavrila96}
M. Marinescu and M. Gavrila, {\em Phys. Rev. A},
{\bf 53}, 2513 (1996).

\bibitem{gavrilabook}
M. Gavrila, Atomic structure and decay in High Frequency fields, in M. Garvila, editor, 
{\em Atoms in intense laser fields}, pages 435-510, (Academic press, New York,
1992).

\bibitem{vorobeichik98}
I. Vorobeichik, R. Lefebvre and N. Moiseyev, {\em Europhys. Lett},
{\bf 41}, 111, (1998).

\bibitem{bagwell92}
P. F. Bagwell and R. K. Lake, {\em Phys. Rev. B},
{\bf 46}, 15329 (1992).

\bibitem{wagner97}
M. Wagner, {\em Phys. Stat. Sol. (b)},
{\bf 204}, 382 (1997).

\bibitem{henseler01},
M. Henseler, T. Dittrich and K. Richter, {\em Phys. Rev. E},
{\bf 64}, 046218 (2001).

\bibitem{georgeot95}
B. Georgeot and R. E. Prange, {\em Phys. Rev. Lett.},
{\bf 74}, 4110 (1995).

\bibitem{magnus54}
W. Magnus, {\em Commun. Pure. Appl. Math.}, {\bf 7}, 649 (1954).

\bibitem{maricq82}
M. Matti Maricq , {\em Phys. Rev. B}, {\bf 25}, 6622 (1982).

\bibitem{salzman87}
W. R. Salzman, {\em Phys. Rev. A}, {\bf 36}, 5074 (1987).

\bibitem{fernandez90}
F. M. Fern\'{a}ndez, {\em Phys. Rev. A}, {\bf 41}, 2311 (1990).

\bibitem{gavbook}
M. Gavrila, editor, {\em Atoms in intense laser fields}, (Academic press, New York, 1992).
 
\bibitem{costin}
O. Costin, J. L. Lebowitz and A. Rokhlenko, {\em J. Phys. A: Math. Gen.},
{\bf 33}, 6311 (2000); O. Costin, R. D. Costin, J. L. Lebowitz and A. Rokhlenko, {\em Commun. Math. Phys.}, {\bf 221}, 1 (2001); A. Rokhlenko, O. Costin and J. L. Lebowitz, {\em J. Phys. A: Math. Gen.}, {\bf 35}, 8943 (2002).


\bibitem{li99}
W. Li and L. E. Reichl, {\em Phys. Rev. B},
{\bf 60}, 15732 (1999).

\bibitem{emman02}
A. Emmanouilidou and L. E. Reichl, {\em Phys. Rev. A},
{\bf 65}, 033405 (2002).

\bibitem{zeldovich67}
Y. B. Zeldovich, {\em Sov. Phys. JETP}, {\bf 24}, 1006 (1967).

\bibitem{shirley65}
J. H. Shirley, {\em Phys. Rev.}, {\bf 138B}, 979 (1965).

\bibitem{sambe73}
H. Sambe, {\em Phys. Rev. A}, {\bf 7}, 2203 (1973).

\bibitem{salzman74}
W. R. Salzman, {\em Phys. Rev. A}, {\bf 10}, 461 (1974).

\bibitem{gesztesy81}
F. Gesztesy and H. Mitter, {\em J. Phys. A: Math. Gen.},
{\bf 14}, L79 (1981).

\bibitem{goldstein}
H. Goldstein, C. Poole and J. Safko, {\em Classical Mechanics},
(Addison Wesley, San Francisco, 2002).

\bibitem{kerner58}
E. H. Kerner, {\em Can. J. Phys.}, {\bf 36}, 371 (1958).

\bibitem{breuer89}
H. P. Breuer and M. Holthaus, {\em Z. Phys. D}, {\bf 11}, 1 (1989).

\bibitem{lefebvre97}
R. Lefebvre and A. Palma, {\em J. Mol. Str. (Theochem)}, {\bf 390}, 23 (1997).

\bibitem{rabi37}
I. I. Rabi, {\em Phys. Rev.}, {\bf 51}, 652 (1937).

\bibitem{tannoudji}
C. Cohen-Tannoudji, B. Diu and F. Laloe, {\em Quantum mechanics} (volume I),
(John Wiley \& sons, New-York, 1977).


\bibitem{barone77}
S. R. Barone, M. A. Narcowich and F. J. Narcowich,
{\em Phys. Rev. A}, {\bf 15}, 1109 (1977).

\bibitem{ido}
I. Gilary, N. Moiseyev, S. Rahav and S. Fishman,
{\em J. Phys. A: Math. Gen.}, {\bf 36}, L409 (2003).

\bibitem{nimrev}
N. Moiseyev, {\em Phys. Rep.}, {\bf 302}, 211 (1998).

\bibitem{averaging}
J. A. Sanders and F. Verhulst, {\em Averaging methods in nonlinear dynamical systems}, (Springer-Verlag, New York, 1985).

\bibitem{scherer95}
W. Scherer, {\em Phys. Rev. lett.}, {\bf 74}, 1495 (1995).

\bibitem{super}
D. Daems, A. Keller, S. Guerin, H. R. Jauslin and O. Atabek,
{\em Phys. Rev. A}, {\bf 67}, 052505 (2003).

\bibitem{messiah}
A. Messiah, {\em Quantum mechanics} (Volume I), (John-Wiley, New-York, 1961).

\bibitem{englman63}
R. Englman and P. Levi, {\em J. Math. Phys.}, {\bf 4}, 105 (1963).

\end{thebibliography}
\end{document}